\documentclass[]{aa}
\usepackage{natbib}
\usepackage{graphicx}
\usepackage{subfigure}
\usepackage{aalongtable}
\usepackage{captcont}
\begin{document}

\title{Age distribution of young clusters and field stars in the SMC\thanks{based on observations taken at ESO under proposal 66B-0615}}
\subtitle{}

\author{E. Chiosi\inst{1}, A. Vallenari \inst{2}, 
  E. V. Held \inst{2}, L. Rizzi\inst{3}, A. Moretti\inst{1} }

   \institute{Astronomy Department, Padova University, Vicolo 
	      dell'Osservatorio 2, 35122 Padova, Italy \\   
   \and   INAF, Padova Observatory,
Vicolo dell'Osservatorio 5, 35122 Padova, Italy   	\\ 
   \and Institute for Astronomy, 2680 Woodlawn Drive, Hawaii 96822-1897, USA \\
   \email{emanuela.chiosi@oapd.inaf.it; antonella.vallenari@oapd.inaf.it;  enrico.held@oapd.inaf.it moretti@pd.astro.it  }               
            }
 
\offprints{A. Vallenari}

\date{Received: November 2005 / Accepted February 2006}

\abstract{}
{In this paper we discuss the cluster and field star formation in the central part of the  Small Magellanic Cloud.
 The main goal is to study  the correlation between young objects and their  interstellar environment.} 
{ The ages of about  164 associations and 311 clusters 
 younger than 1 Gyr are determined using isochrone fitting. The spatial distribution of the clusters is compared with the HI maps, with the HI velocity dispersion field, with the location of the CO clouds and with the distribution of young field stars. } 
{The cluster age distribution supports the idea that clusters formed in the last 1 Gyr of the SMC history  in a roughly continuous way with periods of enhancements. The two super-shells 37A and 304A detected in the HI distribution are clearly visible in the  age distribution of the clusters: an enhancement in the cluster formation rate has taken place from the epoch of the shell formation. A tight  correlation between young clusters and the HI intensity  is found. The degree of correlation is decreasing with the age of the clusters. Clusters older than 300 Myr are located away from the HI peaks.  Clusters and associations younger than 10 Myr are
related to the CO clouds in the SW region of the SMC disk. 
 A positive correlation between the location of the young clusters and the velocity dispersion field of the atomic gas is derived only for the shell 304A, suggesting that the  cloud-cloud collision is probably not the most important mechanism of  cluster formation.
Evidence of gravitational triggered episode due to the most recent close interaction between SMC and \object{LMC} is found both in the cluster and field star distribution.}
{}

\keywords{Galaxies: Magellanic Clouds -- Galaxies:star clusters -- Galaxies: stellar content }

\titlerunning{Cluster and field SFR in the SMC}
\authorrunning{E.Chiosi, A.Vallenari et al.}
\maketitle

\section{Introduction}\label{sec_intro}

 Star formation
is a complex phenomenon  involving at the same time
 several physical  processes such as turbulence, gravitational  
collapse, cooling, gravitational trigger.
 There are many open 
questions about star formation in galaxies:
is it a continuous process or does it proceed by bursts? Which are the fundamental triggers 
of star formation and in which measure are they internal or external?
The Large and Small Magellanic Clouds (\object{LMC}, \object{SMC}) which are believed to be interacting with each other,  are ideal laboratories to study  the process of field star and cluster formation.

  \object{LMC} has been widely studied using both ground based and HST data \citep{bertelli1992,vallenari1996,gallagher1996,holtzman1997,elson1997,geha1998,harris1999,harris2004,olsen1999,dolphin2000,javiel2005}. The vast majority of the authors suggest that the \object{LMC} experimented a continuous SF with several enhancements, roughly at 2-4 Gyr and at 6-7 Gyr although the precise epochs change from field to field. 
At the opposite the \object{SMC} has been less studied.
The \object{SMC} shows an asymmetric appearance with an irregular main body and  an eastern extension. In a photographic plate study \citet{gardiner1992}  found that the bulk of the stellar population in the \object{SMC} is about 10 Gyr old.  They observe that the young stellar population is biased toward the eastern  \object{LMC}-facing side of the \object{SMC}. \citet{crowl2001} find the same trend among the  \object{SMC} populous clusters: those toward the eastern side tend to be younger and  more metal rich than  those on the western side. Recently \citet{zaritsky2000},  \citet{maragoudaki2001}  and  \citet{cioni2000} confirmed that the asymmetric structure of the \object{SMC} is due exclusively  to the distribution of young main sequence stars, while old stars show a  rather regular and smooth distribution typical of a spheroidal body.  The asymmetric distribution of  the young stars is consistent with the  patterns of the associations and HII regions \citep{bica1995}.  This is interpreted as the effect of the perturbations developed by  the interaction of \object{LMC}-\object{SMC}. 

Concerning the SFR of the field stars in the \object{SMC}, no real consensus is reached whether the star formation has proceeded with several  periods of enhancements, namely at 400 Myr, 3 Gyr, 9 Gyr as found by  \citet{harris2004} in the \object{SMC} disk, or in a more continuous way, with a main  episode between 5 and 8 Gyr as derived by \citet{dolphin2001} in the halo.
A large population of clusters is found in the \object{SMC}.
\citet{hodge1986}, comparing the number  of clusters found down to $B$=22 (inner regions) or $B$=23 (outer regions)
in selected regions with the number of known clusters in \object{SMC} catalogs at that time, estimated a global population of 900 clusters. Considering incompleteness effects, 2000 clusters are expected   if small older clusters were detectable. 
Pietrzynski \& Udalski (2000) catalog includes 238 clusters down to $B$$\sim$21.1 or $V$$
\sim$21.5.  \citet{bica2000} including in the catalog faint and loose systems, find 633 clusters. Considering also the clusters related to emission (NC and CN types) the total number of objects is 719.  As pointed out by the authors, the number of clusters in \citet{bica2000} is still far from being complete.
Concerning cluster age distribution,
the \object{SMC} is known  to have  at least six populous  clusters of intermediate age  in the range 5-9 Gyr, but only one true old object  (\object{NGC 121}) having an age of $ > $ 10 Gyr \citep{stryker1985,dolphin2001}. Only a few of the clusters in \citet{bica2000} catalog have known ages.  \citet{pietrzynski1999} using isochrone fitting,  and \citet{rafelski2005} making use  of integrated colors derive the age of a limited number of bright clusters, namely 93, and 200 respectively. \citet{piatti2005} discuss the age and metallicity for 36 \object{SMC} clusters.  The data seem to suggest that the formation  of young clusters took place in bursts \citep{rafelski2005,piatti2001,rich2000}.

The aim of this work is to cast light on the process of field and cluster formation in  this galaxy deriving the age distribution of a large number of  clusters in the central regions of the \object{SMC}  and comparing it with the
field star formation.
At this purpose we make use of the OGLE data for the whole \object{SMC} disk, and of better quality data obtained at the ESO 2.2m telescope  
for a region around \object{NGC~269}, located at the SE end of the disk, at the border of the supershell 37A.                         
In section \S \ref{observation} the observations  and data reduction are described. In section \S\ref{modulus} the \object{SMC} distance, reddening, metallicity, and line of sight depth are given.  
 In section \S \ref{age_det}  the cluster age distribution in the whole disk is derived.
In section \S \ref{clu} the two supershells, 37A and 304A are discussed, 
in section \S \ref{correla}
 the spatial distribution of the clusters is compared with the environment properties. 
In section \S \ref{field_age} field  star formation rate is discussed. Finally, concluding remarks are given
in section \S\ref{conclusions}.

\section {Data, observations, and reduction}\label{observation}
\subsection{\object{NGC~269} region}

$V$,$I$ images were taken with the  WFI at the ESO 2.2m telescope on 2000, October 21 under photometric conditions as backup observations for a different project.
 The field of view is of 34 $\times 33$ arcmin$^2$. It is  centered on the cluster \object{NGC~269} at  $\alpha= 00^h 48 ^m 30.6^s$ and $\delta=-73^0 31^\prime 30^{\prime \prime}$. The exposure times are 300 sec in $V$ and 300 sec in $I$. To avoid saturation at the bright magnitude end two images having exposure times of 20 sec were taken. The seeing was 1 arcsec.
  \begin{figure}

   \centering


    \resizebox{\hsize}{!}{\includegraphics{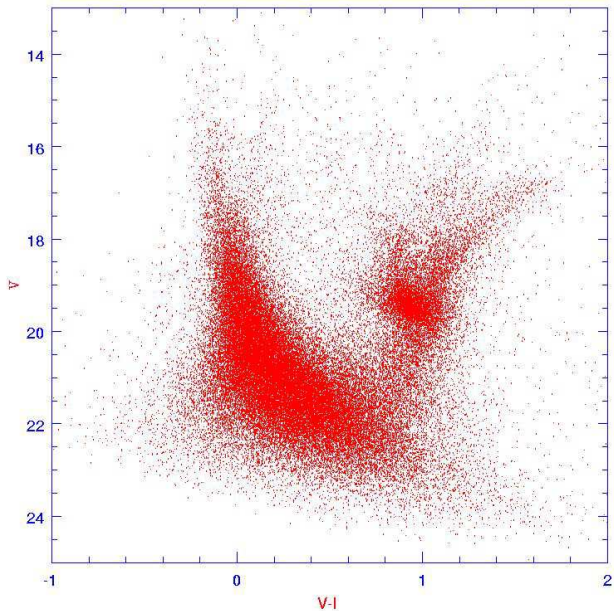}}
    \caption{The  CMD of the observed region around  \object{NGC~269}.}
   \label{cmdtot.fig}
    \end{figure}

\begin{figure}

   \centering

   \resizebox{\hsize}{!}{\includegraphics{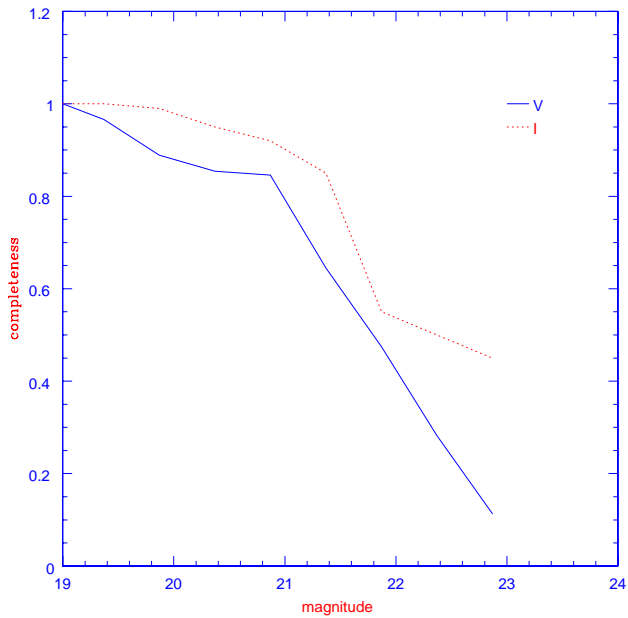}}

    \caption{Completeness factors $\Lambda$ in $V$ and $I$ magnitude are plotted as  functions of the magnitude.}
   \label{comple.fig}
    \end{figure}

\begin{figure}

   \centering

  \resizebox{\hsize}{!}{\includegraphics{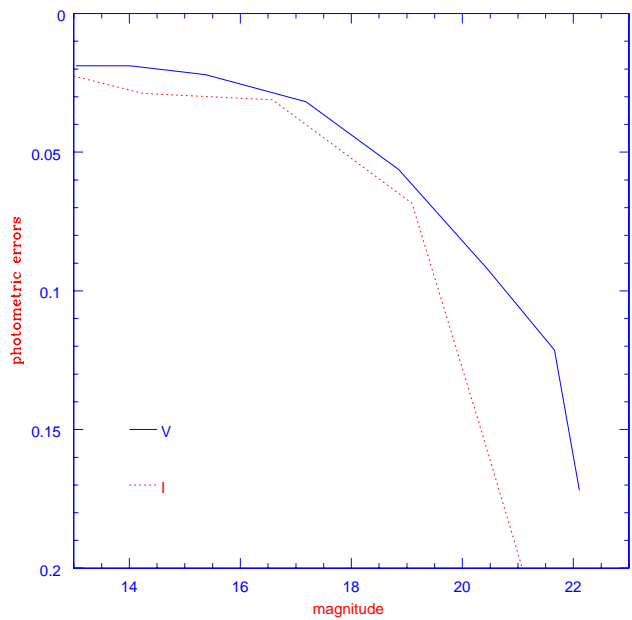}}  

 \caption{Photometric errors as functions of the $V$ and $I$ magnitude.}
   \label{errors}
    \end{figure}

Pre-reduction of CCD images was performed within the IRAF environment. Each image was bias-subtracted and flat-fielded using twilight sky flats. After these steps, all images were astrometrically calibrated using the IRAF package MSCRED (Valdes 1998) and the script package WFPRED developed at the Padova Observatory \citep{Rizzi2003}.
Photometry was obtained  with DAOPHOT (Stetson 1987).The photometric zero points were then set by comparison with secondary standard stars  accurately calibrated onto the Landolt (1992) system. The estimated uncertainty of the zero point calibration is 0.03 mag in both $V$ and $I$.   More than 100000 stars are found down to $V$=24.  Fig \ref{cmdtot.fig} presents the CMD of the whole region.
The completeness correction is calculated as usual by means of artificial stars experiments where a small number of artificial stars are injected in the original frames. Then the frames are reduced following the same procedure. The completeness factors in $V$ and $I$ band, $\Lambda_V$, and  $\Lambda_I$ respectively, defined as the number of recovered on the added stars   are plotted
in Fig. \ref{comple.fig}. The data are complete at 50\% level for magnitudes
brighter than  $V$ $<$ 22. Photometric errors are derived from artificial stars
experiments and are plotted in Fig \ref{errors}.

 \begin{figure}

   \centering

   \resizebox{\hsize}{!}{\includegraphics{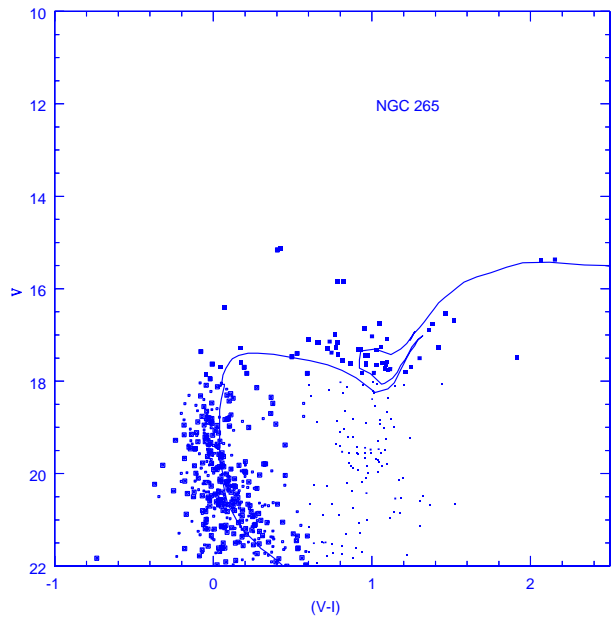}}

   \caption{CM diagram of cluster \object{NGC~265}  showing field stars (light dots) and cluster stars (heavy dots). An isochrone having Z=0.008 and an age of 2.5 $\times 10^8$ is over-plotted on the data.}

   \label{field_sub.fig}
    \end{figure}

\subsection{OGLE data}
An  analysis over a wider area of the galaxy 
is made using 
Optical Gravitational Lensing Experiment (OGLE II) data  \citep{1998AcA....48..147U} to derive cluster and field star age distribution for the  regions not covered by our photometry.
 The data cover 2.4 sq. deg. of the main body of the \object{SMC}.  The limiting magnitude of the photometry is $V$$\sim $21.5.  Completeness correction is applied following \citet{1998AcA....48..147U}.
Field population data are complete at about 80\% level  89\% level down to V=20.5 mag, in the most and in the least crowded fields respectively. For  magnitudes brighter than V=20.5, the uncompleteness correction is not very significant in the field population.  The  50\% level is reached at about $V$=21 mag in the most crowded fields. This mean uncompleteness correction is suitable not only for the field population, but as well for sparse clusters. Inside dense clusters the completeness might be very different.
 In order to estimate the incompleteness correction in these objects, we compare the region of  \object{NGC 269} we observe and of which we estimate the completeness, with OGLE data. We select the core of dense clusters. The incompleteness correction we derive  for the field population is in agreement with the nominal values by  \citet{1998AcA....48..147U}. Inside the radius of the densest clusters given by 
\citet{bica2000}, the OGLE data are complete at 80\%, 70\% and 50\% level  at magnitudes V=19.0,19.5,20.3 respectively. This correction is then applied to derive  the ages of dense clusters.

\section{SMC distance modulus, reddening, line of sight depth, metallicity}
\label{modulus}

A distance modulus of $(m-M)_0=18.9$  is assumed for the SMC, in agreement
with recent determinations  by \citet{storm2004}, \citet{weldrake2004}, \citet{caputo1999} and \citet{sandage1999}.

The extinction map across the \object{SMC} was recently determined by \citet{zaritsky2002} using data from the Magellanic Cloud Photometric surveys. They show that the extinction varies  spatially increasing along the \object{SMC} ridge from  northeast to southwest. In addition, young stars present an average differential extinction 0.3 mag higher than old stars.  
By comparison, a mean reddening 
$E(B-V)$=0.08~mag is derived by \citet{tumlinson2002} and  \citet{hunter2003}. 
In the following, when possible, we derive the reddening of each cluster by  main sequence fitting. Otherwise, a mean value of $E(B-V)$=0.08~mag of the reddening is assumed. In the case of field star population we account for an additional reddening Gaussian dispersion with $\sigma=0.09$. No  differential extinction is included. 

The line of sight depth of the \object{SMC} is a long-lasting controversial issue.
\citet{mathewson1986} derive a great depth of 30 Kpc by measuring distances and radial velocities of Cepheid stars. \citet{welch1987} concluded that the \object{SMC} does not extend beyond its tidal radius (4-9 Kpc). \citet{groenewegen2000} using near infrared data of Cepheids derives a depth of 14 Kpc. 
\citet{crowl2001} using the magnitudes of the clump stars in clusters derive
a depth ranging between 6 and 12 Kpc.
The main reasons of those discrepancies can be found in the uncertainties on the extinction and on the relation period-luminosity-metallicity of Cepheid. \citet{stanimirovic2004} point out that the application of the correction for differential reddening derived by \citet{zaritsky2002}  significantly influences the distance determinations bringing the depth of the \object{SMC} within its tidal radius (4-9 Kpc). Additionally, while the tidal tails contribute mostly to the elongation of the galaxy, the main body of the \object{SMC} does not present a significant elongation ($\sim 5 $ Kpc). A recent determination is made by \citet{2005MNRAS.359L..42L} who using the  P-L relation of OGLE~II variables find a very patchy structure with a depth of about 3.2 $\pm$ 1.6 Kpc.
In the following discussing the SFR of the field stars, we adopt an intermediate value of 4 Kpc implying a difference in the distance modulus
of about 0.14mag.

The present-day knowledge of the age-metallicity relation in the \object{SMC} is mainly based on  clusters. Only a few determinations based on stars are available. RR Lyrae star  abundances are measured by \citet{smith1992}, while Cepheids   metallicity are derived by \citet{harris1981}. The interpretation of the existing \object{SMC} age-metallicity relation widely varies from author to author.  Current data cannot really discriminate among different models. A  continuous enrichment from the oldest to the youngest objects is found  by \citet{dacosta1998} and by \citet{dolphin2001}. \citet{olszewski1996} suggest that no significant enrichment is produced from 10 Gyr ago to 1-2 Gyr ago. At that time the metallicity  rapidly increases. \citet{olszewski1996}, \citet{pagel1999} and  \citet{piatti2001} analyzes of the data favor a bursting mode of star formation \citep{harris2004}. 

In the following,  discussing  the star formation from field stars, we assume the enrichment history by \citet{pagel1999}.

\section{Cluster age distribution}\label{age_det}

In this Section we derive the age distribution of the \object{SMC} clusters located in the main body of the \object{SMC}  using isochrone fitting.
The catalog of the studied objects along with coordinates and radii is taken from 
 \citet{bica2000}.
82 clusters and associations are identified in the region centered on \object{NGC~269}, while 229 clusters   and about 164 associations are studied in the OGLE regions. 
The cluster list, their coordinates, and the derived ages
are available from the authors upon request.
In this Section, we first describe the method, and we compare the age determination with previous studies, then we present the spatial distribution of clusters of different ages.

\subsection{The Method}\label{cmd}

The age and the reddening of each clusters are derived by means of  isochrone fitting on the CMDs in two ways, by visual inspection and by $\chi^2$ minimization. Isochrones are taken from the library of \citet{2002A&A...391..195G}.
The main body of the \object{SMC} is a highly crowded region. For this reason,
field star contamination severely hampers the age determination. 
Field subtraction is a critical issue in order to derive the cluster ages.
When ages are derived by visual inspection, then field stars are statistically subtracted by the CMDs of the clusters. 
First we consider an equivalent area of field close to the 
area of the cluster, but outside the cluster radius given by \citet{bica2000}. Then,
the CMDs of both cluster and field are divided in boxes of size $\Delta V=0.5$ and $\Delta (V-I)=0.2$. The incompleteness correction is taken 
into account by dividing the field and cluster CMDs in magnitude-color bins
and then adding on each bin  having N$_{th}$ stars,
$\Lambda \times N_{th}$ objects, where $\Lambda$ is the smallest of the $V$ and
I completeness factors.  Then, in each box of the cluster CMD, for every field star, the closest cluster  star is subtracted. Finally  isochrones are superimposed on the CMDs in order to fit the location of the main sequence and of the evolved stars.   
When the ages are derived using a $\chi^2$ minimization, first single stellar populations at changing ages are generated using a Monte Carlo method, taking into account the observational errors on the magnitudes.  Then the simulations are corrected for incompleteness, subtracting on each bin  $(1-\Lambda \times N_{th})$ objects. The observational field population corrected for the ratio of the field and cluster incompleteness factors is derived as described in the previous paragraph and  added to the simulated CMDs.
Then the CMDs are divided into  bins of 0.2 both in mag and in color and the $\chi^2$ function of the difference between the observational CMD and the theoretical ones at changing ages is minimized.
A mean metallicity of Z=0.008 is assumed, in agreement with  observational determination for young objects \citep{pagel1999}. However, when the isochrone fitting requires it,  a different metal content is adopted.
Clusters in which the age values derived in both methods are in reasonable agreement are included  in Table \ref{NGC269.tab} and a mean value of the age is given.
 Table \ref{NGC269.tab} gives the catalog of the clusters, their position,  ages, and reddening. 
 Due to the limiting magnitude of the photometry, clusters having turnoff magnitude fainter than $V$=21.5 mag  in our \object{NGC 269} region data, or $V$=20 in OGLE fields  cannot  reasonably be  identified. This sets a limit  of  3 or 1 Gyr, respectively  (assuming Z=0.008), to the oldest age we can derive. For homogeneity, we restrict ourselves to study clusters younger than 1 Gyr.
To minimize the effect on the age determination of the young clusters and associations due to the saturation limit of the OGLE photometry we make use of the bright star catalog and of the catalog of stars of known spectral type by \citet{massey2002}.
To test our method and derive the uncertainties on the age determination, we perform Monte Carlo simulations where synthetic clusters at different ages are generated, field contamination is included and ages are re-derived using $\chi^2$ minimization. Ages derived from integrated colors suffers from several  effects such as discreteness of isochrones, patchy distribution of the interstellar reddening  producing artifacts and spurious peaks in the age distribution \citep{2006MNRAS.366..295D, 2006astro.ph..1606L}. Those effects are less relevant when ages are derived from CMD fitting. Interstellar extinction plays a minor role on the determination of the ages from the main sequence turnoff location. Uncertainties on the interstellar extinction  are of the order of $\Delta(A_V) \sim$ 0.1. In the age range under discussion, this results  in an uncertainty of about 0.03 on log(age). 
The uncertainties on the age determinations are partly a function of the age itself, in the sense that older clusters are more difficultly identified, and partly a function of the cluster density. $\Delta(log(age))$, the errors on log(age) are of the order of 0.22   taking into account the uncertainties on the metal content for relatively dense objects younger than about 2 $\times 10^8$ yr. A poor age resolution is expected for objects younger than 1 $\times 10^7$ yr due to the difficulty of distinguishing massive main sequence and evolved stars, in absence of spectroscopic information. Uncertain membership can further complicate the age determination of those clusters/associations. In addition, the youngest age in the Padova isochrones is log(age(yr))=6.6. Objects younger than this limit are therefore assigned to this minimum age.   Clusters older than  2 $\times 10^8$ yr  have a  mean error $\Delta(log(age)) \sim 0.3$.  This uncertainty is mainly due to the fact  that  Padova isochrones have problems to reproduce both the turnoff and the luminosity of  the clump of He-burning stars in the above age range. Very 
sparse clusters having less than 50 members might have  highly uncertain determinations ($\Delta(log(age) >$ 0.5), especially at old ages.

In Fig. \ref{field_sub.fig} we  present as example the CMD of \object{NGC~265}, one of the bright  clusters  in the region surrounding \object{NGC~269}. The CMD  is fitted with an isochrone having Z=0.008 and an age of 2.5 $\times 10^8$ yr. It is evident that it is difficult to reproduce the color of the main sequence and the location of the red evolved stars at once. This is a well known problem.  Differential reddening can mimic  this effect. However  it cannot be excluded that   the uncertainties affecting the opacities, and/or  the adopted value of the envelope mixing length
are responsible of this discrepancy.
In Table \ref{NGC269.tab} an index gives the degree of reliability of the age measurement we estimate.  Class 1 indicates objects having $\Delta(Log(Age(yr))) < 0.3$; class 2 indicates objects having $0.3 < \Delta(Log(Age(yr))) < 0.5$;  class 3 indicates objects having $\Delta(Log(Age(yr))) >0.5$.

\begin{figure}
   \centering
   \resizebox{\hsize}{!}{\includegraphics{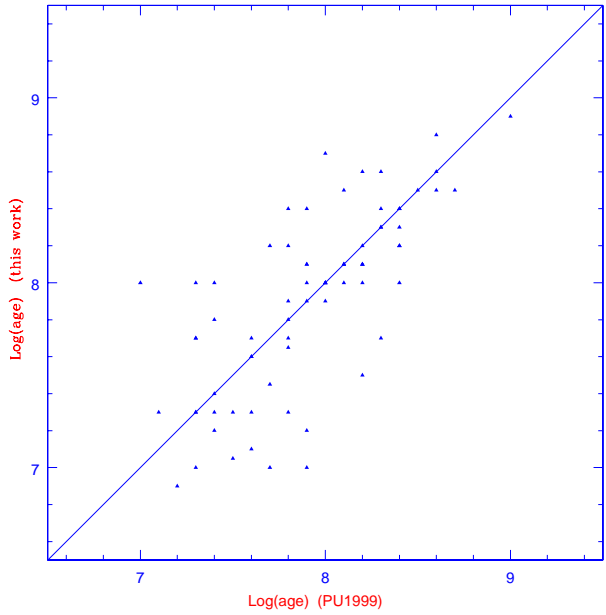}}  

   \caption{Cluster ages derived in this paper are compared with the ages by Pietrzynski \& Udalski (PU1999) for  the clusters in common. The solid line shows the loci of the 1:1 correlation.}

   \label{confronta}
    \end{figure}

\begin{figure}

   \centering

   \resizebox{\hsize}{!}{\includegraphics{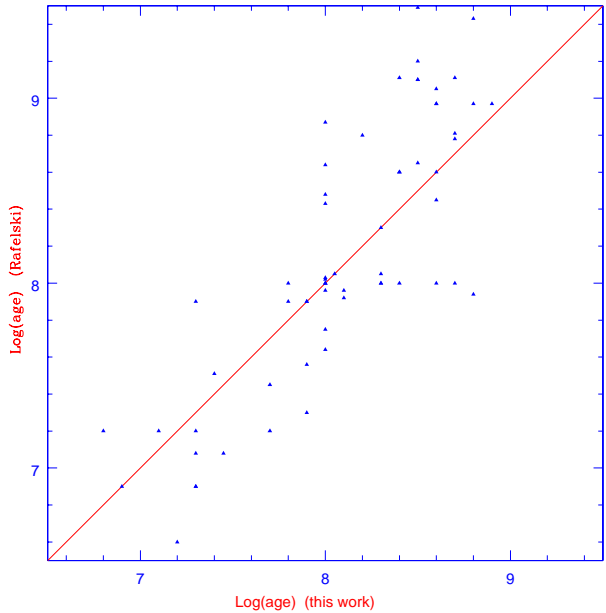}}

   \caption{Cluster ages derived in this paper are compared with the ages by \citet{rafelski2005} (indicated by the label Rafelski) for the clusters in common. The solid line shows the loci of the 1:1 correlation.}

   \label{rafel}
    \end{figure}

In Fig.\ref{confronta} the cluster ages  derived in this paper are compared for the common objects to those obtained via isochrone fitting by  Pietrzynski \& Udalski. Our ages are broadly correlated  to those presented there. No systematic difference is present. The dispersion about the line 1:1 correlation for the whole sample is $\sigma \textrm{log(age)}= 0.3$.
We compare our ages with  the compilation by \citet{rafelski2005} where 204 star clusters are identified and their ages are derived using integrated colors. In principle the color of a stellar population provides a reliable chronometer to date clusters. In practice however, as already pointed out by \citet{rafelski2005} 
 stochastic effects on the number of bright stars, uncertainties on the metallicity and on the adopted stellar models make it difficult to precisely infer  the cluster formation. Fig. \ref{rafel} shows the comparison between ages derived via isochrone fitting  by us and via integrated colors by \citet{rafelski2005}  (for Z=0.004).  
Neglecting a few outliers, the dispersion around the  1:1 correlation line is  $\sigma \textrm{log(age)}/log(age)=0.4$.  A reasonable agreement is found. The outliers having very large uncertainties are all inside large groups of clusters having small separation along the line of sight or they are very sparse objects. In these clusters to derive star memberships on the basis of photometric information is a cumbersome affair.

\subsection{Cluster age distribution in the main body of the \object{SMC}}

\begin{figure}
 \resizebox{\hsize}{!}{\includegraphics{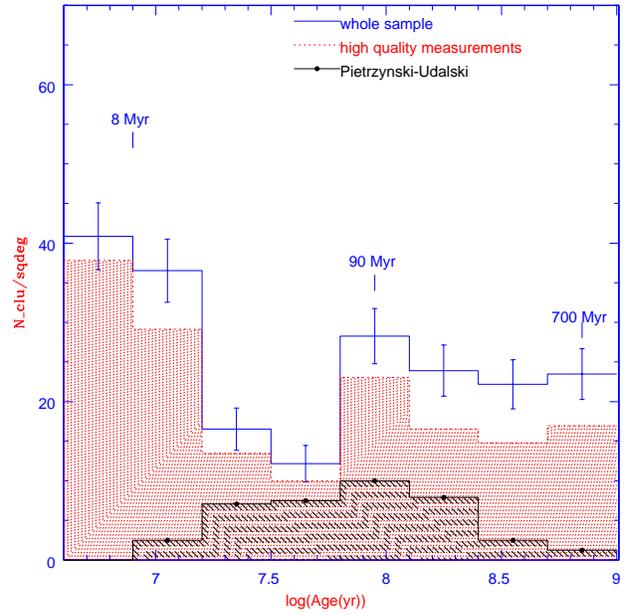}}
   \caption{The cluster age distributions. The continuous line shows the whole sample discussed in this work. The dotted dashed histogram represents the high quality measurements of ages (classes 1 and 2 of Table \ref{NGC269.tab}). Squares indicate \citet{pietrzynski1999}  sample.}
    \label{hist_cfr.fig}
    \end{figure}

   \begin{figure}
   \resizebox{\hsize}{!}{\includegraphics{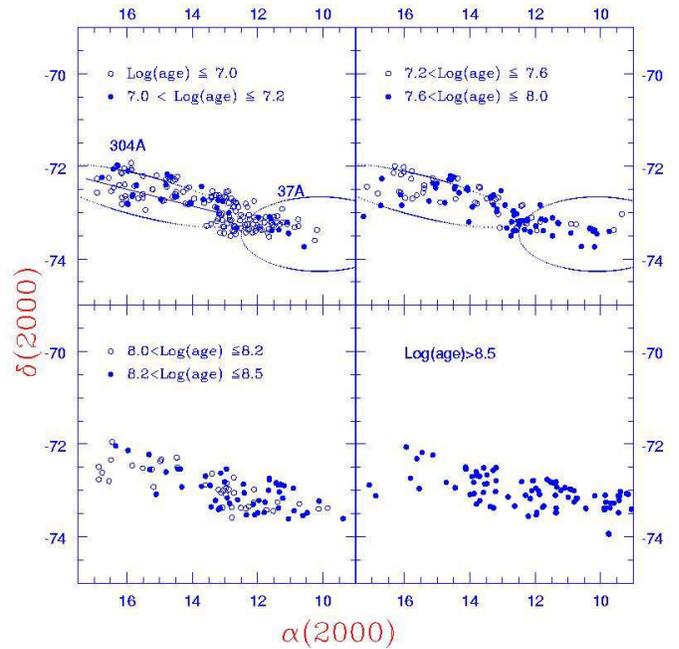}}  
   \caption{ Spatial distribution of clusters of different ages in the \object{SMC}.  The approximate locations of the supershells 37A and 304A are shown.}
    \label{agecfr.fig}%
    \end{figure}

The age distribution of the clusters in the studied regions is shown in Fig. \ref{hist_cfr.fig}. Two main episodes  are found, the first from 5 to 15 Myr, the second  at 90 Myr about. This latter is followed by an almost constant rate till 1 Gyr. The uncertainties on the age determinations do not allow a finer age resolution.
We point out that these peaks are present in the whole sample and in the  clusters having the most reliable age measurements (classes 1 and 2 in Table \ref{NGC269.tab}).     
Looking at the spatial distribution of the clusters (see Fig. \ref{agecfr.fig}), 
it is evident that clusters in a given age range are not uniformly distributed across the SMC, but the bursts are associated with localized enhanced SFR. Since rotational mixing did not had enough time to smooth the cluster distribution, this points in favor of the fact that these SFR enhancements  are not due to artifacts in the age determination procedure, but are likely  real. 

\begin{figure}
   \resizebox{\hsize}{!}{\includegraphics{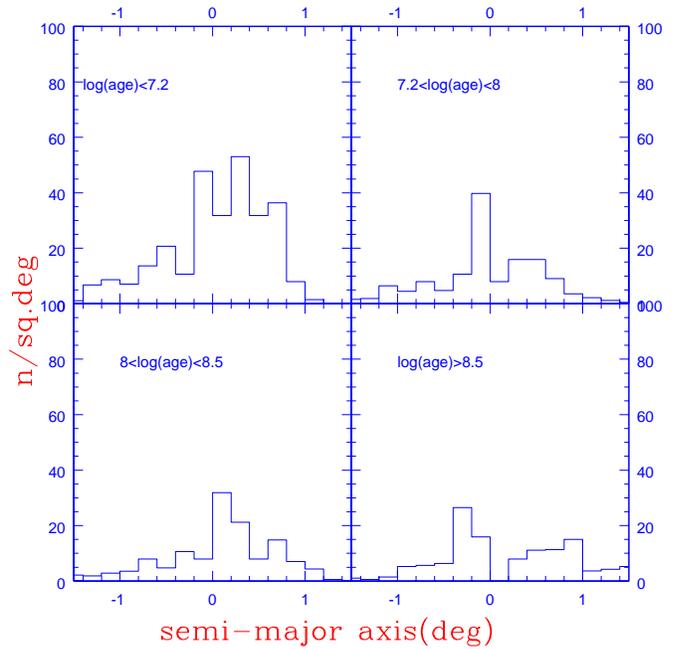}}  
   \caption{ Cluster distribution as a function of distance from the \object{SMC} center  at changing ages. Negative axis values indicate the regions East of the center, while positive axis points toward the West.}
    \label{smc_bar}%
    \end{figure}

To describe the spatial distribution of the clusters, we model the \object{SMC} disk as an ellipse 
centered at  $\alpha= $  00$^{\rm h}$ 52$^{\rm m}$ 45$^{\rm s}$, $\delta= $-72 $\hbox {$^\circ $ }$ 49 $\hbox{$^\prime$ }$ 43 $\hbox{$^{\prime\prime}$ }$ (J2000) (Crowl et al. 2001)
and having an axial ratio  $b/a=1/2$. Then we define the distance along the major axis as the major axis of the ellipse  having the same center and axial ratio and passing on the object. 
 Fig.\ref{smc_bar} presents the  distribution of the clusters per area unit at changing ages as a function of the distance from the SMC center. 
A  complex picture is emerging.
Two main regions located East and West of the center were active at very young ages ($<15$ Myr), while the cluster formation process was less significant 100 Myr ago. These regions can be identified with the two HI super-shells 37A and 304A. 
More detail can be found in the following sections, where the two regions will be analyzed. 

\begin{figure}
   \resizebox{\hsize}{!}{\includegraphics{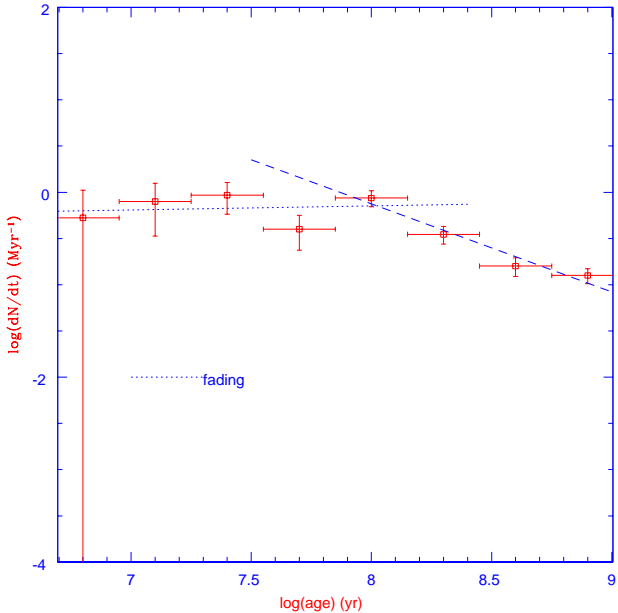}}  
   \caption{ Cluster age distribution(squares). Only objects classified as C are included in the sample. The data are fitted by two lines. The dotted line represents  the fading line, while the dashed line is the fit for ages where most likely the disruption and the sample incompleteness dominate fading.}
    \label{disruption}
    \end{figure}

Finally, we address the question whether the age distribution we find is  representative of the cluster formation rate or whether the tidal field of the SMC was  effective in  disrupting the less massive clusters.
  Following the discussion by  \citet{boutloukos2003}, in a survey of clusters having a given magnitude limit, two effects contribute to the
the age distribution of clusters born at a time $t_0$. The first is the fading. Clusters get fainter with time as a result of the evolution of their stars.  As a result, the number of observable clusters as a function of age for a given magnitude limit is decreasing. This effect is dominant for young objects. 
The second is the cluster disruption due to the galactic tidal field and is  relevant for old clusters. A steep slope of  $dN_{obs}/dt$ is expected. 
 The  mass of a clusters decreases almost linearly with time, until the cluster is finally disrupted.  This defines the disruption time, $t_{dis}$. Ignoring burst, the mean cluster formation rate might be assumed as roughly constant. Under this hypothesis, a slope change in the distribution $dN_{obs}/dt$ is expected at a time $t_{cross}$ where the effects of the disruption begin to be significant.  $t_{cross}$ depends on the photometric evolution of the stellar populations, but is as well a function of the magnitude limit of the cluster sample.
Fig.\ref{disruption} presents $dN_{obs}/dt$, the observed age distribution of the \object{SMC} clusters. Only objects classified as type C by \citet{bica2000} are included in the sample.
The age distribution of the \object{SMC} clusters is rather flat for objects with ages below log(t)=8.0 and decreases steeply at higher ages. This identifies  $t_{cross}$. It cannot be excluded that the sample of clusters we discuss is biased toward dense and massive objects which are more easily recognized than less dense or lower mass objects. However  the fact we find a flat distribution at young ages  supports the idea that selection effects, fading and incompleteness play a minor role for clusters younger than this limit. In principle $t_{dis}$ can be derived from $t_{cross}$ and the slope of the disruption line. However in our case, the fact that the completeness correction of the cluster sample is substantially unknown prevents any determination. In fact the sample of ages we derive  is probably  not complete especially at old ages. 
We remind that the 
\object{SMC} cluster disruption time derived in literature is  of the order of 
$8 \times 10^9$ yr \citep{delafuentemarcos1997,boutloukos2003}.

\section {Triggering mechanism of cluster formation}
\label{clu}

In this Section, we would like to address the complex relation between the clusters and their environment, in order to cast light on  the cluster formation process.
 Several mechanisms of cluster formation have been proposed in literature. Here we quote a few.
\object{SMC} is believed to have been involved in  several  perigalactic encounters with the Milky Way and with the \object{LMC}, over the past 12 Gyr.
The epoch of the most recent perigalactic encounter with the \object{LMC} is found to range from Ã¢ÂÂ¼500 Myr ago \citep{lin1995} to 200 \citep{gardiner1994,yoshizawa2003}.  At that time the star formation is expected to be enhanced not only in the tidal arms, but also in the main body of the \object{SMC}.
Clusters are expected to form  as a result of relatively high velocity cloud-cloud collisions \citep{zhang2001, bekki2004}. This mechanism is particularly efficient during galaxy interactions and mergers. The value of the  turbulent velocity between clouds that can give rise to cluster formation is a highly uncertain parameter, going from 50-100 Km/s \citep{zhang2001} to 20-30 Km/s  \citep{bekki2004}. 
Alternatively, high speed motions may produce a high-pressure environment that in turn, can trigger turbulence or shocks \citep{elmegreen1997}. Finally, star formation might be triggered by stellar winds and supernovae 
explosions through compression by turbulent motions (Larson 1993).

The correlation, if any, between young star clusters and their environment can cast light on the cluster formation process.
In the following, we first discuss the age distribution in the regions of  the two super-shells. Then  we discuss the degree of clustering of the clusters and young field stars.
Finally, we compare the spatial distribution with the column density of the gas observed in the HI  and with the dispersion velocity field.

\subsection {Super-shells and cluster formation}

The interstellar medium of the \object{SMC}  shows a fractal structure, consisting of a hierarchy of HI clouds and shells. Kinematic studies of the HI data have revealed the presence of two super-shells in the \object{SMC} disk, namely 37A and 304A \citep{stanimirovic1999} which may consist of smaller superimposed shells. 
The true origin of the holes and small shells in the interstellar medium is still under discussion.  There is evidence that at least  10\% of the small shells found in the \object{SMC} are not associated with young star formation. For this reason, it is unlikely that those few shells are the result of supernova explosions \citep{hatzidimitriou2005} but they might be due to turbulence and gravitational instabilities \citep{dib2005}. However  the vast majority of the shells and supershells is associated with young objects and it is probably   formed in the standard way, because of the combined effects of supernovae and stellar winds \citep{mccray1987}.
Inside the shells, sequential or secondary star formation is expected to be triggered by supernova explosions.  In the following we will discuss the spatial distribution of the clusters/associations of different ages inside  37A and 304A to clarify the relation between cluster formation and super-shells.

\subsubsection{Cluster age distribution in the region of the supershell 37A}
\label{37A}

\begin{figure}
   \centering	    
   \resizebox{\hsize}{!}{\includegraphics{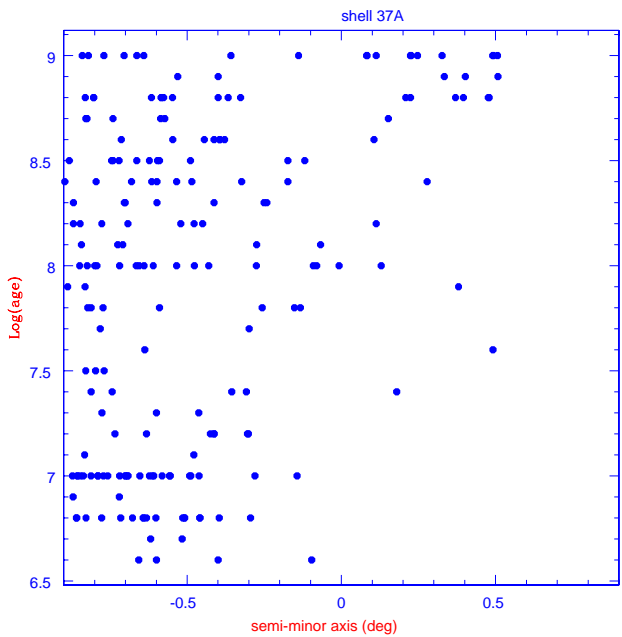}}  
\caption{Age distribution inside the supershell 37A plotted against the semi-minor axis (see text for details). Negative axis refers to the  Eastern part of the shell.}

 \label{age_axismin37A.fig}
    \end{figure}

In order to bring into evidence the spatial distribution of the clusters of different ages, we plot the age distribution of the clusters against the distance from the center. As we did in the previous Section, we approximate the shell with an ellipse whose center and axial ratio are given by \cite{stanimirovic1999}.
For sake of clarity we remind that  the supershell is centered at $\alpha= 00^h 40^m 26^s$ and $\delta= -73^0 28^\prime 06^{\prime \prime}$ and has a major axis a=840 pc (or 0.8 deg), axial ratio of 0.89 and a position angle of $160^0$  \citep{stanimirovic1999}.
 Because of its position angle, the minor axes of the ellipses are roughly aligned in the direction E-W. Fig. \ref{age_axismin37A.fig} presents the cluster age distribution as a function of the semi-minor axis.  Our  data do not completely cover the supershell at the Western side, but the vast majority of it is included in the data. While objects older than  about 15 Myr are found East and West of the center, a discontinuity in the spatial distribution of younger clusters is present. They are almost all  located toward  the Eastern  rim of the supershell 37A where gas and dust are located \citep{staveley1997, stanimirovic1999}. Only two clusters as old as 15 Myr are
found West of the center.  
This discontinuity clearly indicates the epoch of the shell formation which was preceded by a period of relative quiescence. This determination of the supershell age is in agreement with the dynamical age of  17 Myr \citep{stanimirovic1999}.

 It is quite difficult to ascertain the presence of secondary or propagating star formation events
 as  it is expected in the standard model by \citet{mccray1987} if stellar winds and supernovae explosions were responsible for the formation of the supershell.  Standard model of shell formation predicts that young objects are located at the edges, while older stars are more centrally concentrated.
The fact that objects younger than the dynamical estimate of the supershell age
 are located toward the Eastern rim of the supershell, while older clusters are widely distributed might be interpreted as a mild age gradient. 
 However, we remind that the  
 whole analysis is complicated by the extension of the \object{SMC} along the line of sight.  The apparent distance of the clusters from the center might be due to projection effects.

\begin{figure}
   \centering	    
\resizebox{\hsize}{!}{\includegraphics{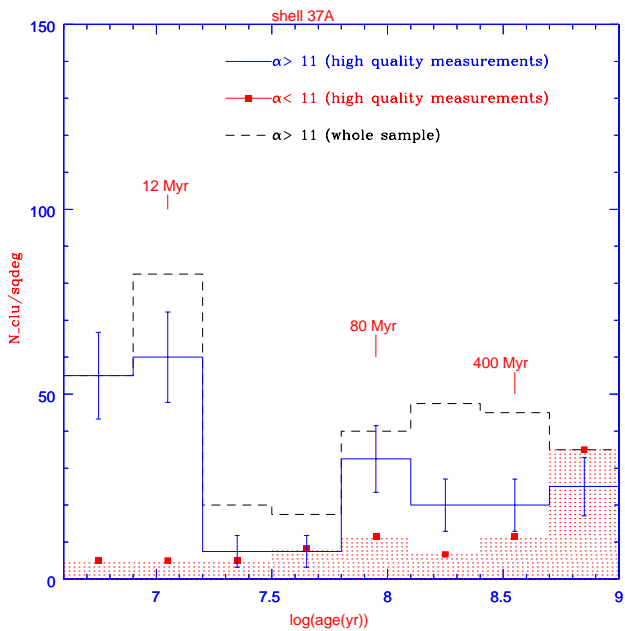}}
\caption{ Dashed line is the age distribution of the whole cluster sample on the Eastern side of the shell 37A ($\alpha > 11$ deg or $\alpha<00^h 44^m 00^s$). Continuous line is the analogous for high quality measurements (see text for details).  Squares show the cluster age distribution on the Western side of the shell 37A ($\alpha < 11$ deg ) for clusters having  high quality measurements of the age.  Error bars indicate the Poissonian uncertainties on the counts.}  
         
    \label{histo_shell.fig}
    \end{figure}

A quantitative description of the cluster age distribution is shown in  Fig. \ref{histo_shell.fig}. We subdivide the supershell region in two parts, one West, from  $\alpha=00^h 44^m 00^s$ (or 11 degrees) to $\alpha=00^h 36^m 00^s$ (or 9 deg.) and one West of $\alpha=00^h 44^m 00^s$. 
The age distribution   presents two episodes that might have different origin. 
  Looking at the cluster age distribution, (see  Fig. \ref{histo_shell.fig}) it is evident that the epoch of the shell formation is coincident with an enhancement in the cluster formation rate in the Eastern part.

In addition to this young episode,  the cluster age distribution at the Eastern part shows an enhancements between 80 and 400 Myr.   On the Western side the cluster formation was less efficient at ages younger than 15 Myr, while it was comparable at older ages.

\subsubsection {Cluster age distribution in the region of the supershell 304A}
\label{304A}

\begin{figure}[t]
\parbox{8.7cm}{
\resizebox{8.7cm}{!}{\includegraphics{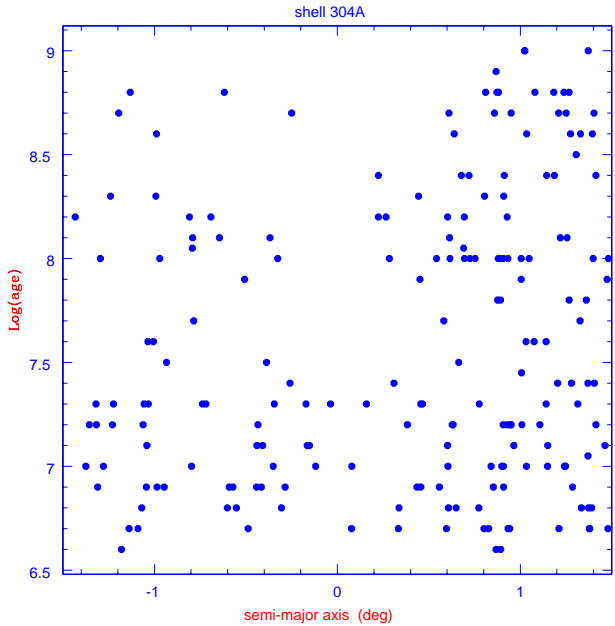}}\\
}

\caption{Age distribution inside the supershell 304A plotted against the semi-major axis (see text for details). Negative values of the semi-major-axis indicate objects located East of the center, while positive values refer to clusters West of the center.
}
\label{age_axismax304.ps}
\end{figure}

\begin{figure}
   \centering	    
\resizebox{\hsize}{!}{\includegraphics{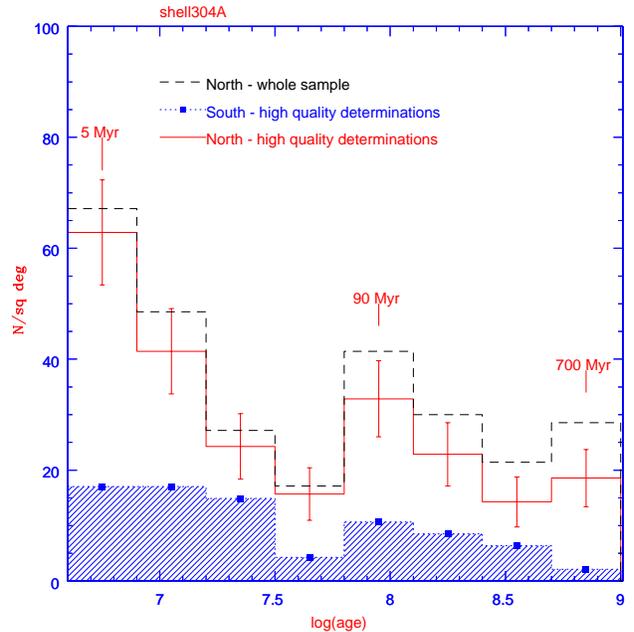}}
\caption{Dashed  line shows the cluster age distribution for the whole sample of objects inside  the shell 304A North of the line drawn in Fig. \ref{agecfr.fig} where the large regions of H$_\alpha$ emission are located. Continuous line is the analogous for clusters having high quality determinations of the age,   while dashed histogram is the analogous at the Southern side. Error bars indicate the Poissonian uncertainties on the counts.   
          }
    \label{shell304A_isto.ps}
    \end{figure}

Analogously to what we did in previous Section, we approximate the supershell region with an ellipse and discuss the distribution of the clusters inside homologous ellipses having the same center and axial ratio of the shell. We remind that  the supershell is centered at $\alpha= 01^h 02^m 16^{s}$ and $\delta= -72^0 38^\prime 12^{\prime \prime}$ and has a major axis a=910 pc (or 0.87 $^0$), axial ratio of 0.45 and a position angle of $80^0$  \citep{stanimirovic1999}. 
Fig. \ref{age_axismax304.ps} shows the cluster distribution as a function of the distance from the center calculated along the major axis of the ellipse representing the supershell (see previous section).
The distribution of the clusters along the major axis (approximately oriented in the direction W-E) clearly shows that at any ages the star formation  took place preferentially in the Western part of the shell, in the direction of the shell 37A.  An almost symmetric distribution is visible for ages younger than 20 Myr, while older objects are mainly located West of the center. This epoch roughly indicates the time of the formation of the shell and is in agreement with the dynamical age of the supershell of 14 Myr \citep{stanimirovic1999}.   
A quantitative description of the age distribution is derived 
subdividing the region in two parts defined by the line drawn in Fig.\ref{agecfr.fig}, roughly separating the Northern region from the Southern.  Fig. \ref{shell304A_isto.ps} presents the age distribution of the clusters. 
The star formation was more active in the Northern region,  where  the majority of the H$_\alpha$ emission is located.
The cluster distribution in the supershell 304A region  shows a continuous formation from a few Myr to 1 Gyr with  enhancements  from a few Myr to 15 Myr, and at 90.  The youngest episode is 
 in coincidence with the epoch of the formation of the supershell.  
 
Summarizing this section, the two super-shells are clearly visible in the age distribution of the clusters. From the epochs of their formation up to now, an enhancement in the cluster formation took place. It is evident that the same mechanism (SN explosions, stellar winds, turbulence) producing the shells in the gas distribution is responsible of the formation of the objects younger than 15-20 Myr. The super-shells  probably formed in more complex way than what is described by the simple model by \citet{mccray1987}:  even if young objects are more numerous at the edges of the shells, a clear age gradient from the centers to the rims is not evident. The inter-shell region   was specially active, possibly due to compression phenomena related to the expansion of the shells.  Finally, an episode at about 90 Myr is found in both shells, even if it is more relevant in the shell~304A. This latter is possibly related to the most recent epoch of  close interaction between \object{SMC} and \object{LMC}.

\section {Correlation of the clusters with their environment}
\label{correla}

In this section we first make use of the correlation function to discuss the degree of clustering of the objects and derive the spatial scale of the formation.
Then we explore the correlation between the clusters and their environment, in particular the HI  flux and velocity dispersion, and the location of the CO clouds.

We adopt the two-point correlation function as a description of  the distribution of clusters and associations with the HI flux map by \citet{stanimirovic1999}. This allows a quantitative measurement of the clustering to complement the visual exploration of the maps (see following).
 $\xi(r)$, the autocorrelation function, is  defined using  the probability $1+\xi(r)$ of finding a neighbor in a shell element of volume $d^3r$ at a distance r from an object of the sample as:
 
$$1+\xi(r)= 1/(Nn) \sum_{i=1}^N n_i(r)$$

where $n_i(r)$ is the number density of objects found in an annulus centered on the i-th object and having radius between r and r+dr, N is the total number of objects, and finally, $n$ is the average number \citep{peebles1980}. A Monte Carlo algorithm is used to derive the area included in the data when the annulus extends outside the studied region.

When the clusters are associated with a continuous map, then the cross correlation function  
is defined as:

$$1+\xi(r)= 1/(Nf) \sum_{i=1}^N f_i(r)$$   

where $f_i(r)$ is the average flux in an annulus with radius r centered on the object, and f is the average flux over the whole region.
Using the above definitions, a random distribution of clusters will produce a flat correlation, with $\xi(r) \sim 0$.  A peaked $\xi(r)$ at small radii  indicates a positive correlation, and the full width half maximum of the peak itself represents the spatial scale of the association between the cluster distribution and the flux. The absolute value of $\xi(r)$ is a measure of the concentration of the flux at a given distance r from the cluster center, relative to the average flux.
Since only clusters in the disk of the \object{SMC} are under discussion, the regions of the HI map outside the OGLE fields are masked out.

\subsection{ Clustering of star clusters and field stars}

\begin{figure}
   \centering

   \resizebox{\hsize}{!}{\includegraphics{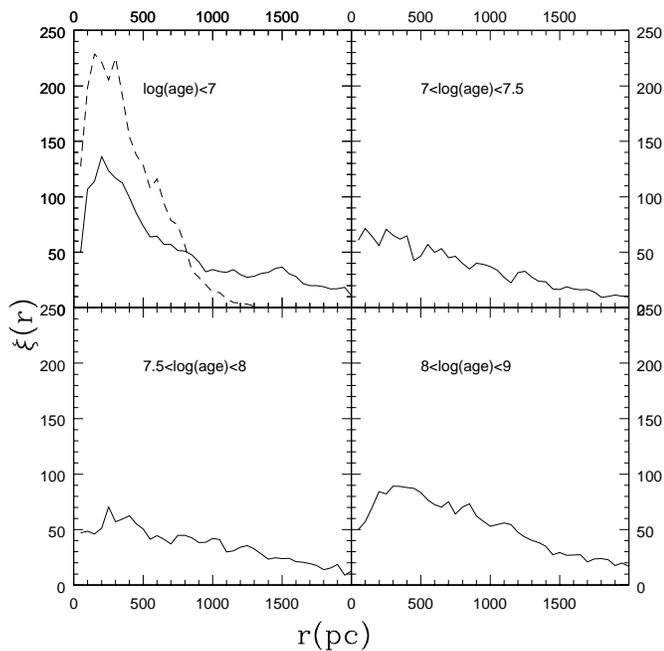}}  
   \caption{ Autocorrelation function of clusters(solid line) and field stars younger than 10 Myr (dotted line). }

    \label{autocross}
    \end{figure}

Fig. \ref{autocross} presents the autocorrelation function of the clusters.
Objects younger  than 10 Myr show a peaked distribution with half maximum full width  of the order of 500pc. 
This distribution  might reflect the structure of the interstellar medium from which they formed. The scale of the clustering is larger than the typical size of molecular clouds in the \object{SMC} and \object{LMC} which is going from about 10 to 100 pc \citep{israel2003} but is comparable  to the size of the complexes and groups of molecular clouds found in the \object{LMC} and \object{SMC} \citep{mizuno2001, israel2003}. The autocorrelation function of the clusters is getting flatter  with age, implying a weaker correlation. In fact, because of cinematic effects, older clusters are more spread out than young objects.
In order to verify whether clusters and field stars show evidence of  different formation mode, we compare the distribution of the young clusters and young field stars in the \object{SMC} disk using the bright star survey by \citet{massey2002}.
We select only field stars younger than 10 Myr  having $V$ $<$ 14. 
Fig. \ref{autocross} presents the autocorrelation function of the bright field stars showing the same spatial  scale as the clusters, but a higher degree of correlation.

\subsection {Cross-correlation of the cluster distribution  with HI map}

\begin{figure*}[t]
\hspace{-0.2cm}
\parbox{9.8cm}{
\resizebox{9.8cm}{!}{\includegraphics{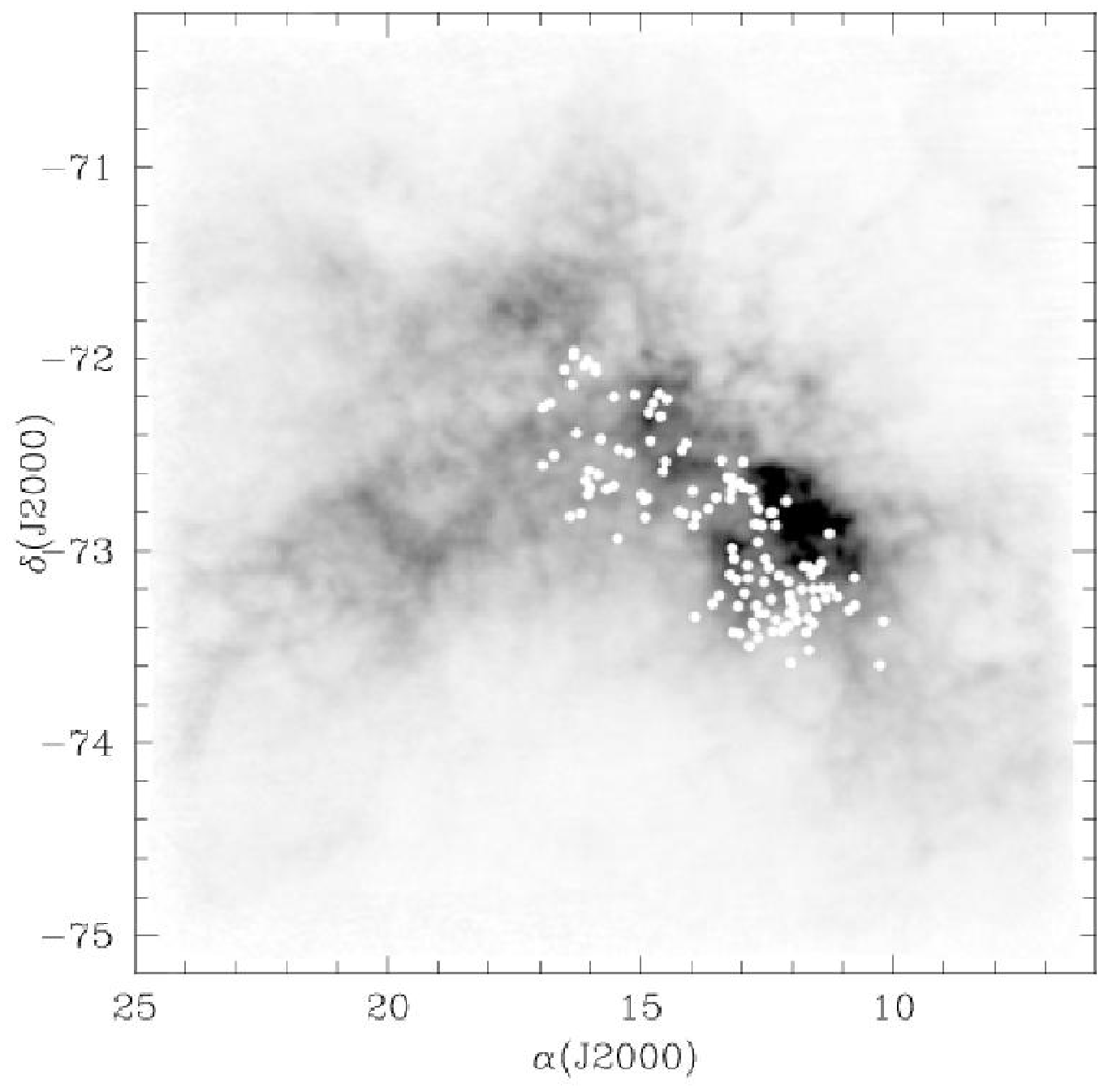}}\\
}
\parbox{9.8cm}{
\resizebox{9.8cm}{!}{\includegraphics{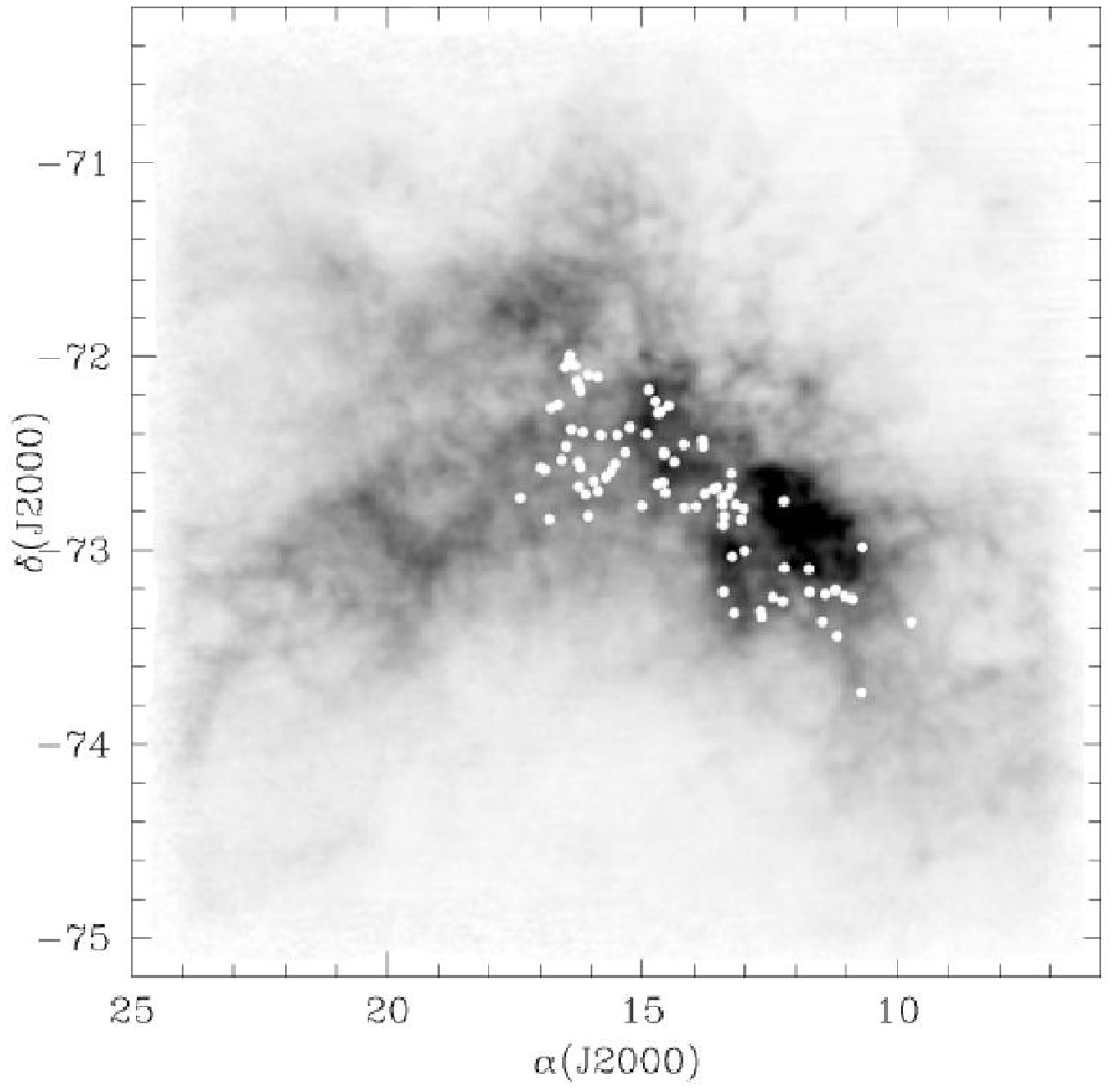}}\\
}
\parbox{9.8cm}{
\resizebox{9.8cm}{!}{\includegraphics{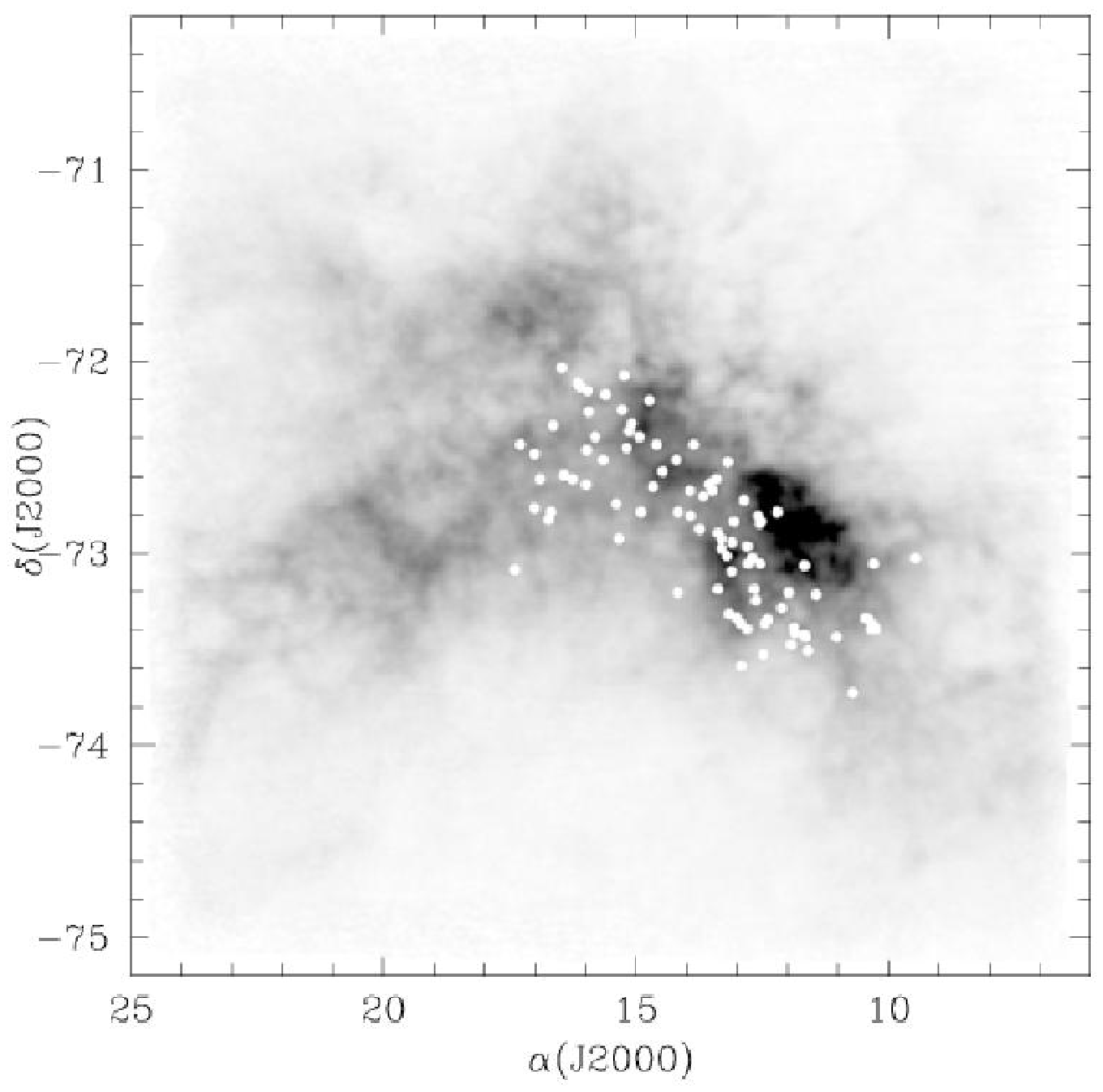}}\\
}
\parbox{9.8cm}{
\resizebox{9.8cm}{!}{\includegraphics{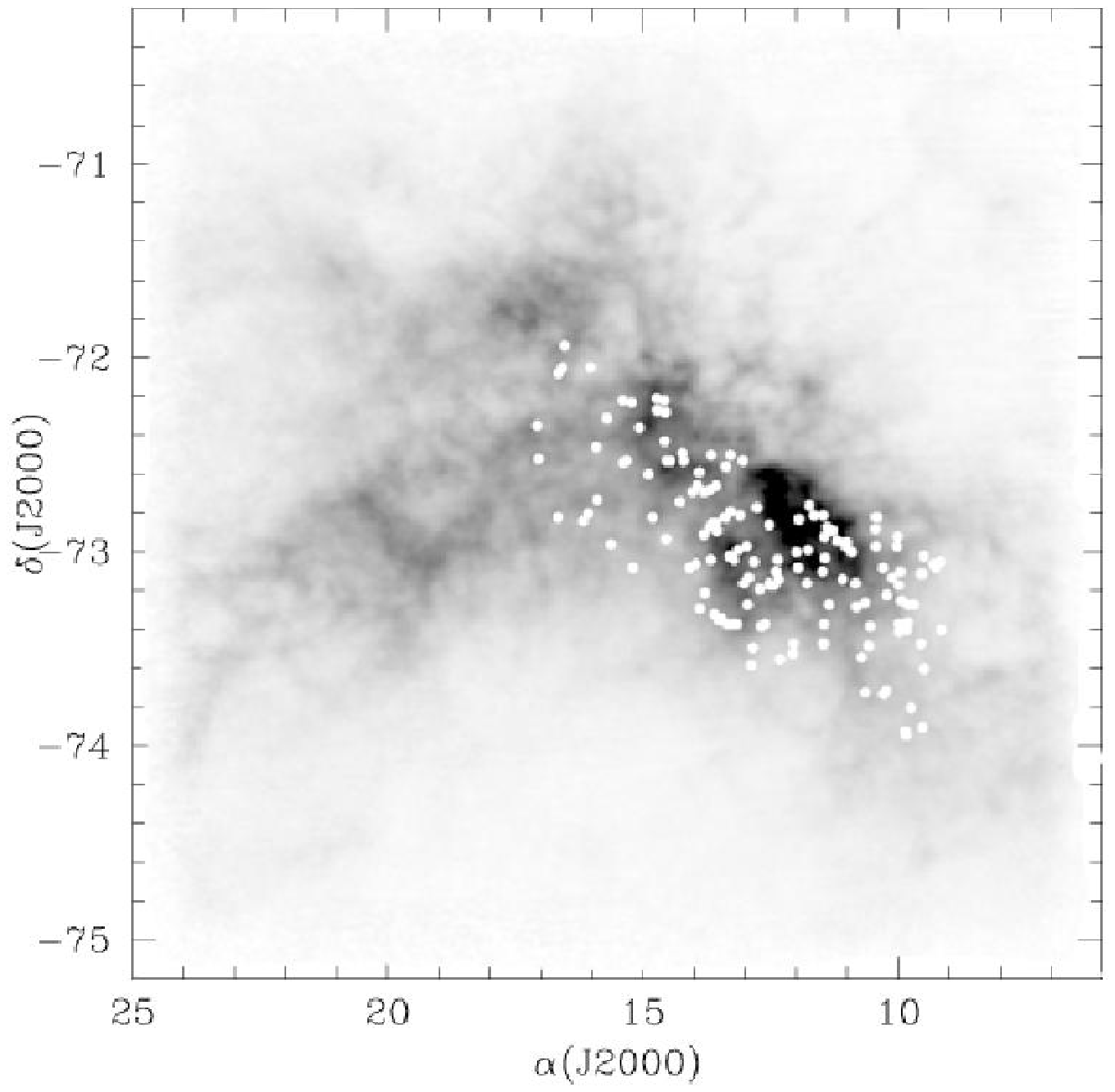}}\\
}

   \caption{
The HI column intensity map in  \object{SMC} is compared with the location of clusters of different ages. Top left panel refers to object younger than 10 Myr; top right panel presents clusters in the age range 10-30 Myr; bottom left panel shows the objects having ages going from 30 to 100 Myr; bottom right presents  clusters from 100 to 1000 Myr old }

\label{fig_Hi_clu}
 \end{figure*}

\begin{figure}
   \centering
    \resizebox{\hsize}{!}{\includegraphics{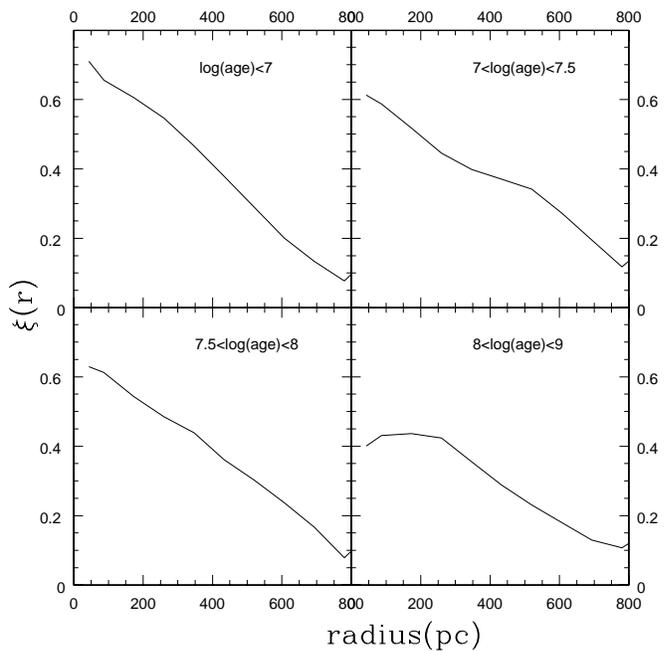}}  
    \caption{Correlation function between  clusters and HI flux map. 
 }
    \label{Hi_correl}
    \end{figure}

The HI 21-cm emission line provides a measurement of the content and kinematics of the atomic hydrogen.
  Fig.\ref{fig_Hi_clu} compares the HI map by \citet{stanimirovic1999} with  the location of the young clusters. 
The region of the maximum HI intensity is located at $\alpha= 00^h 47^m 33^s$ and $\delta= -73^0 05^{\prime} 26^{\prime \prime}$ and shows cluster formation at the edges, where the rim of the supershell 37A is located.
A quantitative measurement of the degree of correlation between clusters and HI is presented in  
Fig.\ref{Hi_correl} showing  the correlation function  at different ages.
Clusters younger than 10 Myr show a positive correlation with the HI map. The degree of correlation is decreasing with age. The correlation is the weakest for clusters in the age range 300-1000 Myr, with a plateau in $\xi(r)$ in the inner  250 pc, implying that the clusters are located away from the peaks of the HI. 
 This positive correlation confirms that the formation of clusters is related to the presence of atomic gas. 

\subsection{Comparing cluster distribution and HI velocity dispersion}

\begin{figure*}[t]
\parbox{9.8cm}{
\resizebox{9.8cm}{!}{\includegraphics{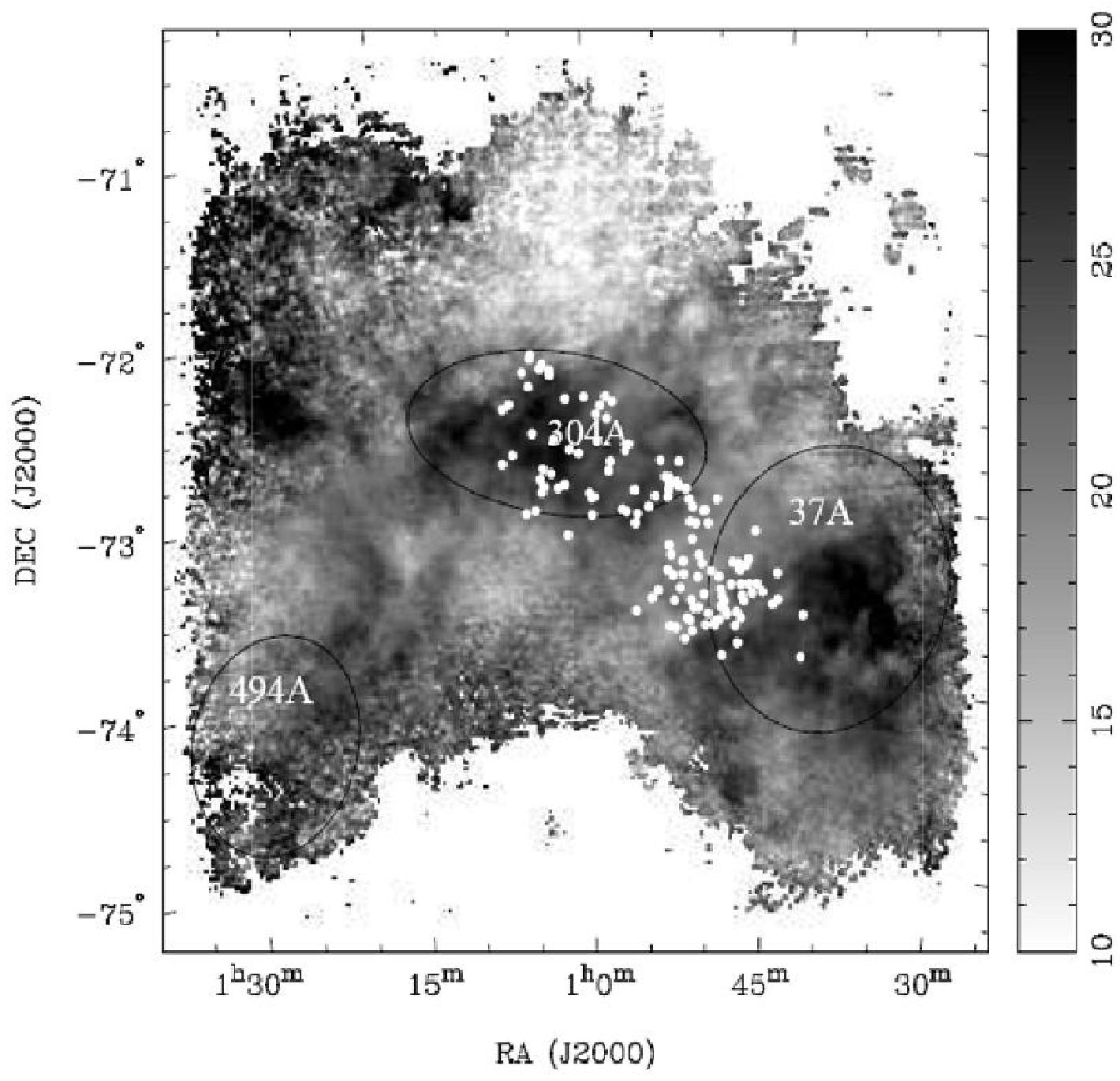}}\\
}
\hspace{0.2cm}
\parbox{9.8cm}{
\resizebox{9.8cm}{!}{\includegraphics{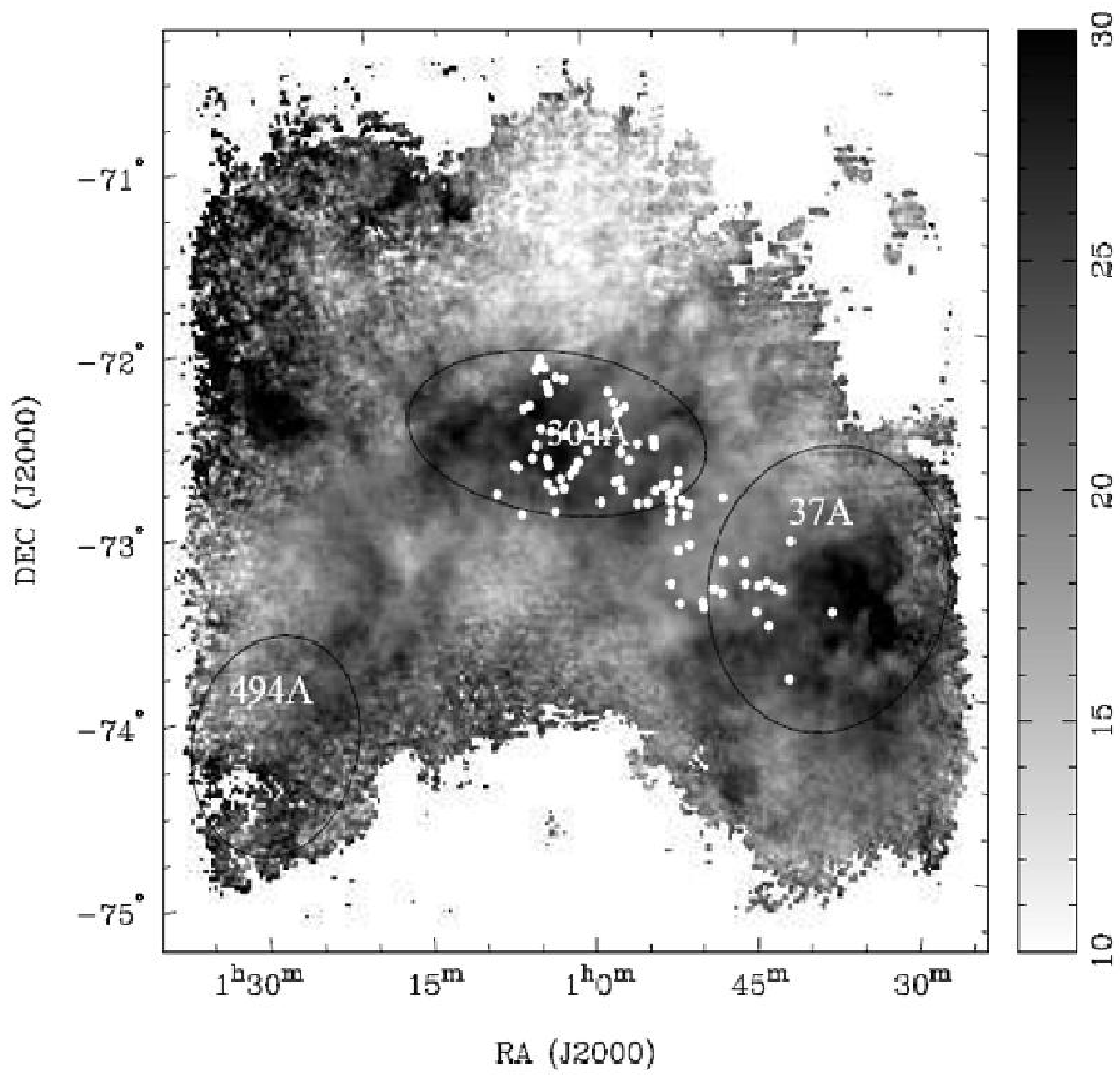}}\\
}
\parbox{9.8cm}{
\resizebox{9.8cm}{!}{\includegraphics{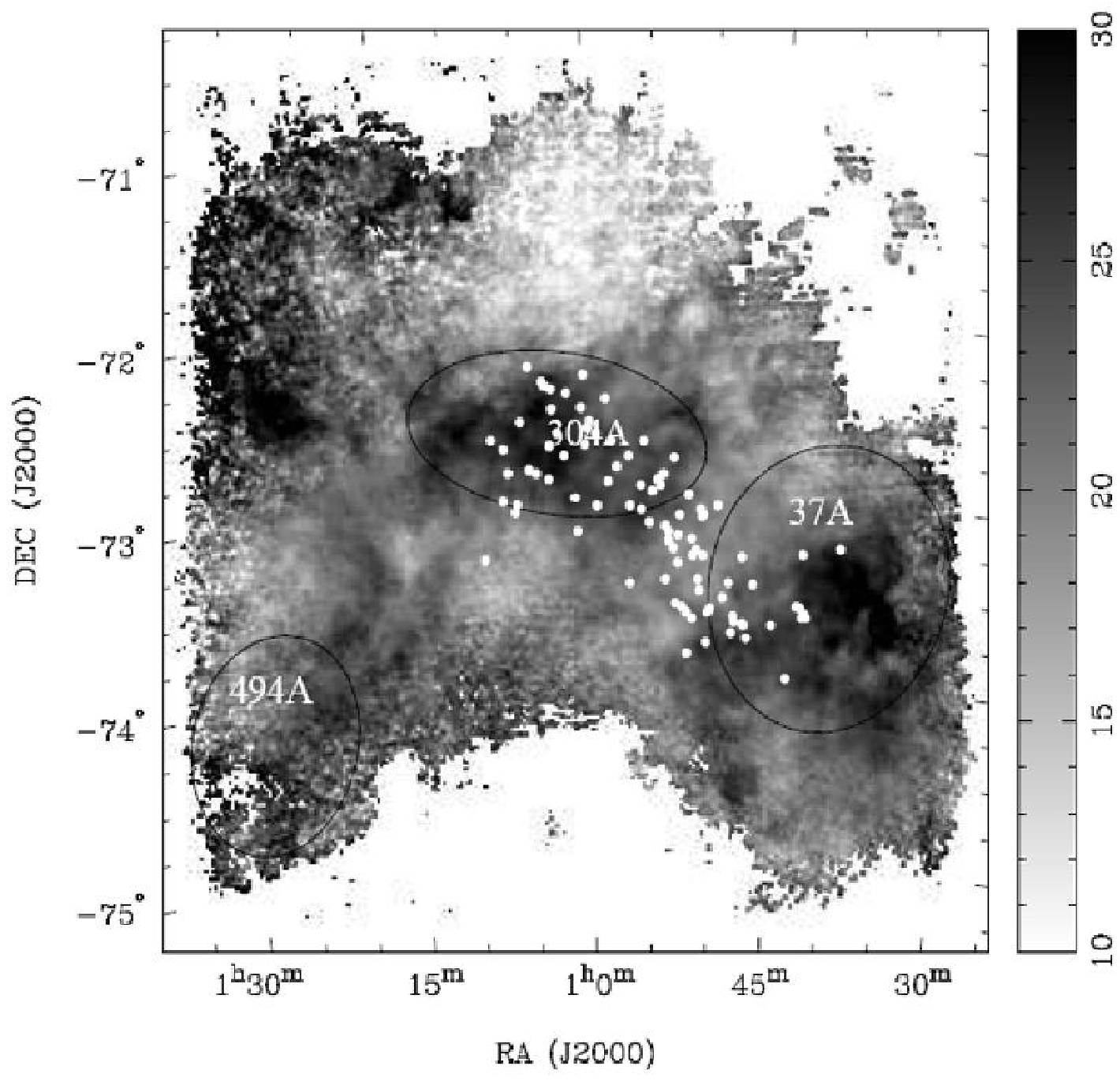}}\\
}
\hspace{0.2cm}
\parbox{9.8cm}{
\resizebox{9.8cm}{!}{\includegraphics{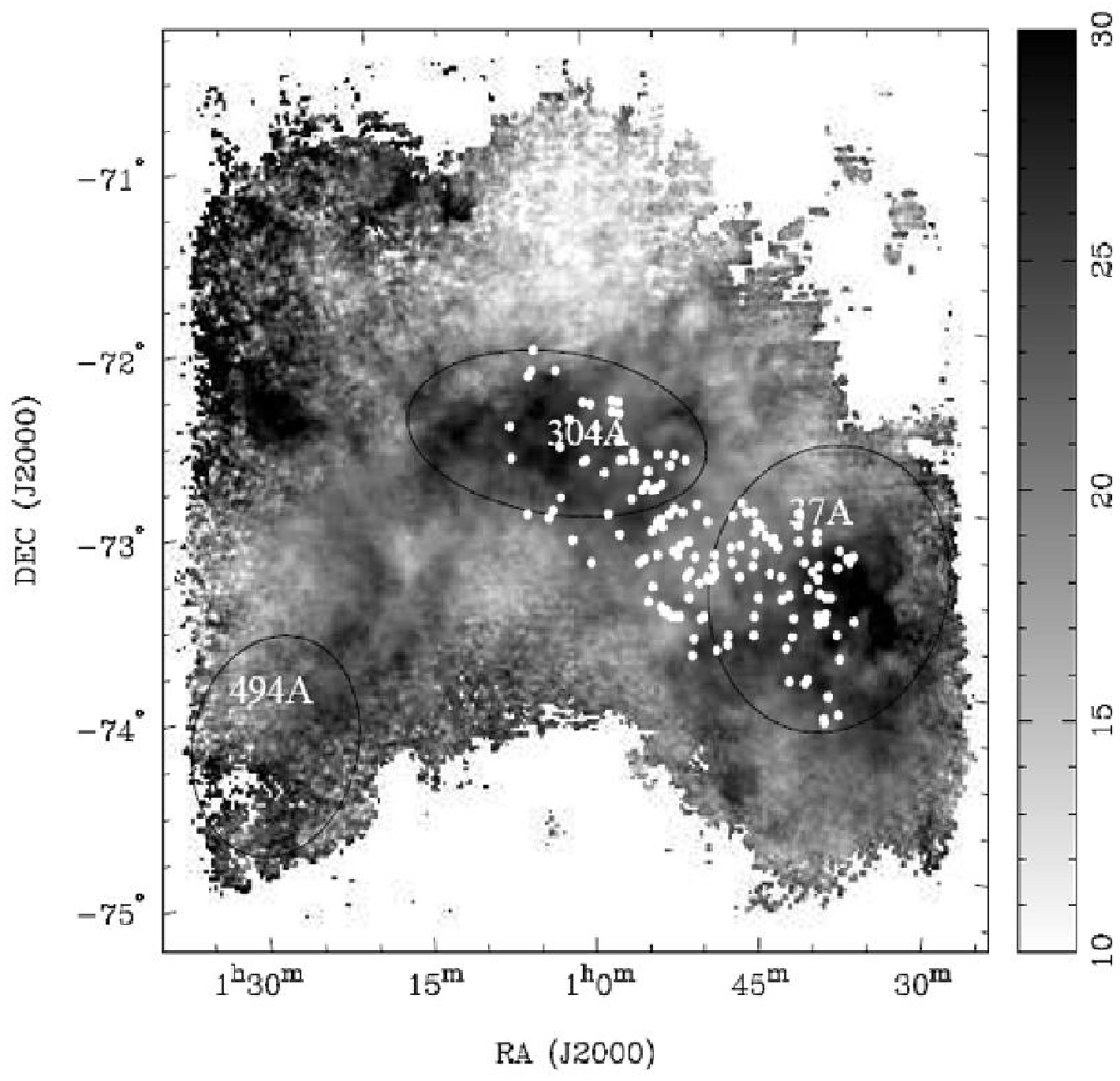}}\\
}

   \caption{
The velocity dispersion field of the HI in  \object{SMC} taken by Stanimirovic et al 2004 is compared with the location of clusters of different ages. Top left panel  refers to object younger than 10 Myr; top right panel  presents clusters in the age range 10-30 Myr; bottom left  panel  shows the objects having ages going from 30 to 100 Myr; bottom right panel  presents  clusters from 100 to 1000 Myr old. }
\label{fig_vel_clu}
 \end{figure*}

A comparison between the cluster distribution and the velocity field provides additional information about the formation and evolution of young clusters. 
Tidal interaction between the \object{SMC} and \object{LMC}/MW involves a  motion of the gas, which in turn might cause shocks and trigger star and cluster formation. 
The quasi periodic approaches of the \object{SMC} to the \object{LMC} sustain tidally and hydrodynamically the large scale turbulent motions in the gas \citep{kumai1993}. 
In fact the \object{SMC} gas is found to present large scale motion elongated on the line of sight \citep{mathewson1986, stanimirovic1999}.
On the other hand, supernovae explosion and  stellar winds from young clusters can put energy into the interstellar medium influencing the  motion of the gas. So a correlation between clusters and high velocity dispersion of HI can cast light on both the cause and the effects of the cluster formation.
We compare in Fig.\ref{fig_vel_clu} the velocity dispersion field of the HI in  \object{SMC}
\citep{stanimirovic2004} with the location of the clusters of different ages. If clusters are formed in high velocity gas motion, then we expect a correlation between the position of young objects and the velocity dispersion of HI.
The super-shells 37A and 304A are clearly visible as disturbances on the velocity field. In the shell 304A about 48\% 
of the clusters younger than 
10 Myr are apparently located in the high dispersion regions ($\sigma_v > 25$ Km/s).
The percentage of objects related to high velocity motion is still  45\% in the age range 20-100 Myr, and is about 35\% at ages of 1 Gyr.
In the shell 37A clusters and associations younger than 10 Myr  are found in regions of intermediate velocity ($\sigma_v \sim 15-25 $ Km/s). Only the 8\% of young clusters is apparently located  in the high velocity regions.  Surprisingly,  about 30\% of clusters in the age range 100-1000 Myr seems to be correlated with high velocity motions, which might be due to chance superposition along the line of sight. If we assume that the distribution of the clusters is uniform, then the probability of a chance superposition is given by the ratio of the area of the high velocity regions over the total disk area.
Under this assumption, the probability of a chance superposition is of the order of 32\%.

Summarizing this section, cloud collisions might have triggered cluster formation in the region of 304A. However this mechanism seems not to account for the formation of clusters in 37A.

\subsection {CO clouds}

Finally, to derive more information  about the recent SF process,
we compare  the distribution of the clusters and of the CO clouds.
CO emission provides a probe of molecular gas inside the galaxies, since it is related to H$_2$ column density.
\citet{rubio1993}, \citet{mizuno2001}  reveal the presence of two main complexes of CO molecular clouds
located in the South West  $\alpha=0^h 45^m$ and $\delta=-73^0 35^\prime$  and North East at $\alpha=1^h 00^m$ and $\delta=-72^0 20^\prime$. They find that the location of
Giant Molecular Clouds (GMCs) shows a good spatial correlation with the HII regions 
and young emission  clusters  indicating that cluster formation is ongoing.  

\begin{figure}
   \centering

   \resizebox{\hsize}{!}{\includegraphics{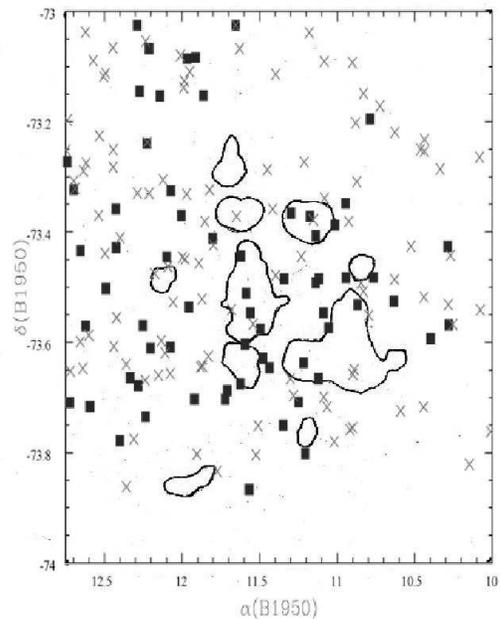}}  
   \caption{ The distribution of clusters and  associations younger than 10 Myr (squares), and
  clusters older than 10 Myr  (crosses)  is compared with the approximate location and size  of the  CO clouds (heavy solid line) 
 from \cite{mizuno2001} in the region of the shell 37A. }

    \label{map_mean.fig}
    \end{figure}

\begin{figure}
   \centering

   \resizebox{\hsize}{!}{\includegraphics{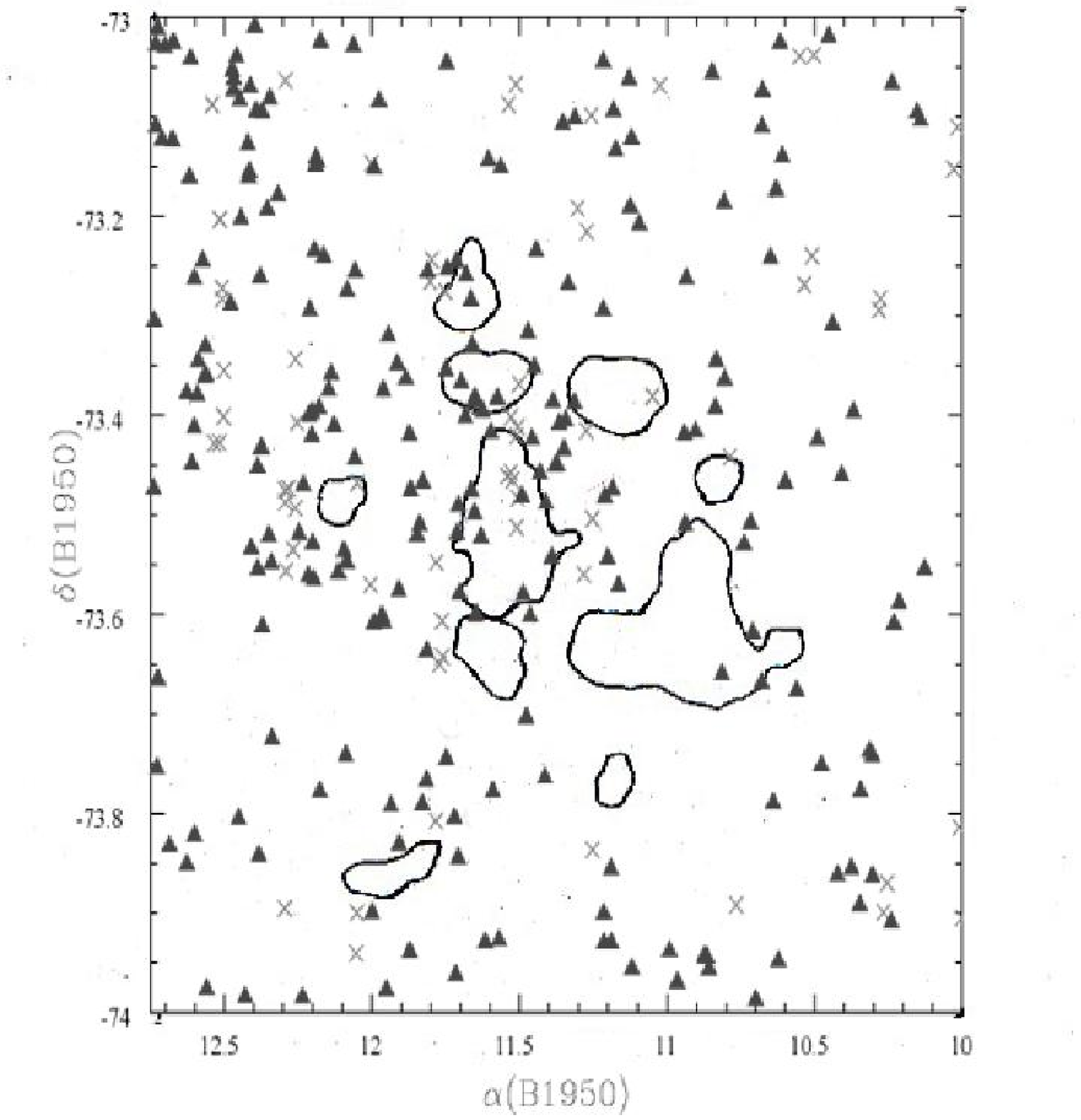}}  
   \caption{ The distribution of field stars younger than 6 Myr (crosses),  and
  stars in the age range 6-10 Myr   (triangles)  is compared with the approximate location and size  of the  CO clouds (heavy solid line) 
 from \cite{mizuno2001} in the region of the shell 37A. }

    \label{map_mean_field.fig}
    \end{figure}

\begin{figure}
   \centering

   \resizebox{\hsize}{!}{\includegraphics{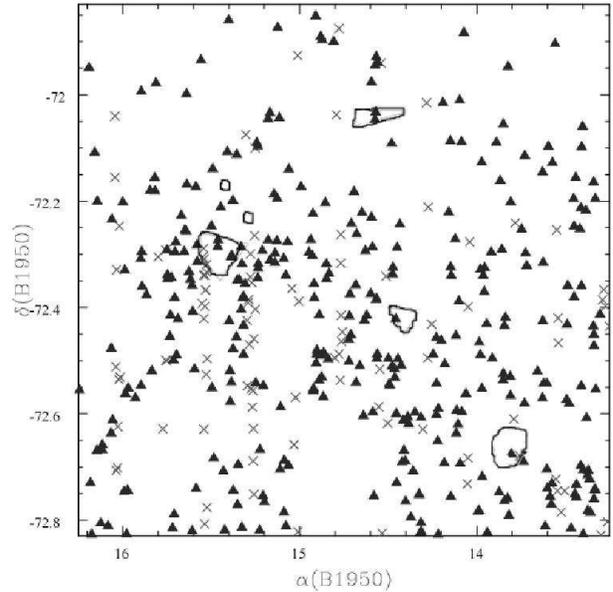}}  
   \caption{ The distribution of field stars younger than 6 Myr (crosses),  and
  stars in the age range 6-10 Myr  (triangles)  is compared with the approximate location and size  of the  CO clouds (heavy solid line) 
 from \cite{mizuno2001} in the Northern region of the disk. }

    \label{map_mean_field_north.fig}
    \end{figure}

We compare the age distribution with the CO cloud catalog by \citet{mizuno2001}
in the SW region of the disk. This region is located at the Eastern rim of the supershell 37A.
We find that about 20\% of the clusters older than 10 Myr are  located close to  the CO
 emission (Fig.\ref{map_mean.fig}). The percentage of associated clusters becomes  35\% when   younger objects are considered.  Assuming a uniform cluster distribution, the probability of a chance superposition is of the order of 11\% \citep{mizuno2001}. A significant fraction of the objects younger than 10 Myr are probably physically associated with the clouds. 
Our result  implies a rapid dissipation of the CO clouds.
This is in agreement with the evolutionary time scale of the giant molecular clouds derived in the \object{LMC} by Fukui (2005). 
 Fig.\ref{map_mean_field.fig} compares the location of the clouds with  field stars
younger than 10 Myr (or brighter than $V$=14) taken from the catalog by \citet{massey2002}.
The 38\% of the stars younger than 6 Myr (or brighter than $V$=13) are associated with the clouds, while only 25\% of the stars  in the age range 6-10 Myr are located in the proximity of the CO emission. 
This gives further support to the idea that the formation of young clusters and field stars is closely related to the CO clouds.
In addition, about 70\% of the stars younger than 10 Myr and associated with the clouds are located close to the Eastern side of the clouds themselves.
We point out that the SW CO clouds are located toward the interface between the two super-shells, 37A and 304 A: dynamical  effects due to the expansion of the super-giant shells may be important in triggering the compression of the molecular clouds and the formation of new stars. Fig.\ref{map_mean_field.fig} might indicates that young stars are more easily found at the compression edge of the clouds. This effect is more relevant for field stars, while is not obvious for the cluster distribution.
The Northern region studied by \citet{mizuno2001} where CO clouds are found   is not included in OGLE data.  
 Fig.\ref{map_mean_field_north.fig} shows the location of field stars brighter than $V$=14 and of the CO clouds in this region. The data are from \citet{massey2002}. Only a negligible percentage of stars is apparently associated to the CO clouds.
This is in agreement with \citet{mizuno2001} who find  a substantial difference between the CO cloud complexes in the SW and in the North of the disk. In the Northern region, the clouds are smaller in size than those of SW region.  No particular correlation is found with the distribution of young associations and clusters except for the objects associated with N66.
In our Galaxy massive stars are formed in clusters in giant molecular cloud (GMC) complexes and almost no GMCs are found in the solar vicinity lacking of massive young stars \citep{fukui2005}.
Discussing the distribution of CO clouds in the \object{LMC}, \citet{fukui1999} and  \citet{fukui2005} find a different behavior, classifying the giant molecular clouds in three groups: 1) without massive OB stars (which does not exclude that low mass stars are formed); 2) associated with small HII regions;
3) associated with clusters and large HII regions. About 38\% of the molecular clouds in the \object{LMC} are not related to recent massive star formation.
 If  CO emission surveys over larger areas  confirm this finding, then the properties of massive star formation regions are probably  different in the Milky Way from those in the MCs.

\section{Field star formation}
\label{field_age}

In this Section we derive the star formation rate of the field stars in order to compare it with the  cluster formation epochs. The goal is to verify whether field objects and clusters follow similar modes of formation.
 We make use of our data inside the 
supershell 37A and of OGLE data in the rest of the disk.
In the following, we first present the method, then we discuss the star formation history of the field population in the two super-shells, comparing the periods of enhancements with what we derive for the clusters.

\subsection {The Method}

To infer the SFR of this galaxy,
theoretical CMDs in different age ranges are simulated.  Simulations include
the spread due to observational photometric errors and reddening as described in the previous Sections. 
For each age bin, from 10000 to 15000 stars were generated down to the photometric completeness limit.
The generation of the synthetic populations makes use of the
set of stellar tracks by \citet{girardi1996,girardi2000}.
The constancy of the initial mass function (IMF) slope in different environments is still a matter of 
discussion, although a number of recent papers  
proposed the idea of a universal IMF \citep{kroupa2002,wyse2002,chabrier2003,weidner2004}. The determination of the IMF is beyond the scope of this paper.
Here the IMF of 
Kroupa (2001, 2002) is assumed.
This IMF is a power-law function with a slope $x=2.3$ for stellar
masses m $> 0.5$ M$_{\sun}$, while $x=1.3$ in the mass range 0.08-0.5 M$_{\sun}$ 
(when the standard Salpeter value is 2.35).

The completeness of the data is taken 
into account by dividing the simulated CMD in magnitude-color bins
and then subtracting from each bin  having N$_{th}$ stars,
(1-$\Lambda$)N$_{th}$, where $\Lambda$ is the smallest of the $V$ and
I completeness factors as given in Fig.\ref{comple.fig}. 

\begin{figure}
   \centering	    
   \resizebox{\hsize}{!}{\includegraphics{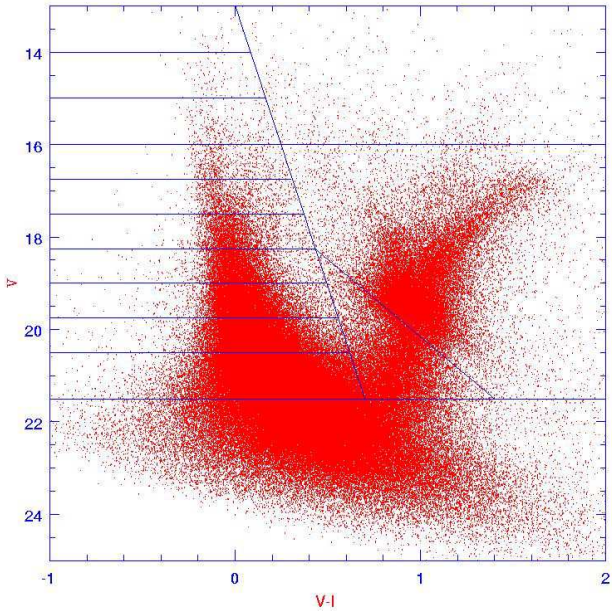}}

\caption{ Bin division of the CMD}
    \label{grid.fig}
    \end{figure}

Finally, the SFR is derived by means of a
downhill simplex method \citep{harris2004},
minimizing the $\chi^2$ function in a 
parameter-space having N dimensions.
At each step the local $\chi^2$ gradient
is derived and a step in the direction of the gradient is taken, till a
minimum is found. 
In the following,
the observational CMD is divided into  bins. Recent work concerning the determination of the
SFR from the CMDs has pointed out the importance of using a
binning, that takes into account the various stellar evolutionary 
phases, as well as the uncertainties on the stellar models \citep{2002ASPC..274..490R}.
In section \S \ref{age_det} we show that problems arise when isochrones are used to fit the main sequence and the He-burning stars.
For this reason, while a coarser magnitude bin distribution is used on the main sequence,
only a few bins are considered for the red evolved stars.
This avoids that the uncertainties on both the observational errors
and the theoretical models (i.e. on bolometric corrections, 
RGB and AGB location, extension of the core He-burning loop) 
result in spurious solutions. For this reason, we are not very sensitive to the SFR at ages older than 1 Gyr.
This is illustrated in
Fig. \ref {grid.fig}.
To prevent settling on local rather than global minima, the simplex is first
 started
from a random position, then when a possible solution is obtained,
it is re-started from a position very close to it. Finally, when a  minimum
is found  30000 random directions are searched for a new minimum.

The first guess solution is obtained by comparing the observational CMD with 
isochrones of different ages and metallicities.
We use N=16 stellar populations, whose ages, metal content
are listed in  Table \ref{pop.tab}.

\begin{table}
\caption{Synthetic population age and metallicity.}
\begin{center}
\begin{tabular}{ c  c }
\hline
\hline
Age  & Z  \\
\hline
5e6:1.5e7 &  0.006 \\
1.5e7:2.5e7 & 0.006 \\
2.5e7:4.0e7 & 0.006 \\
4.0e7:6.3e7 & 0.006 \\
6.3e7:1.0e8 & 0.006 \\
1.0e8:1.6e8 & 0.006 \\
1.6e8:2.5e8 & 0.006 \\
2.5e8:4.0e8 & 0.006 \\
4.0e8:6.3e8 & 0.006 \\
6.3e8:1.0e8 & 0.006 \\
1.0e9:1.6e9 & 0.001:0.006 \\
1.6e9:2.5e9 & 0.001:0.003 \\
2.5e9:4.0e9 & 0.001:0.003 \\
4.0e9:6.0e9 & 0.001:0.003 \\
6.0e9:10.0e9  & 0.001:0.001 \\
10e9:12e9    & 0.001:0.001 \\
\hline
\end{tabular}
\end{center}

\label{pop.tab}
\end{table}

\subsection{Field star formation history  in the HI-shells}

\begin{figure}
   \centering	    
\resizebox{8.9cm}{!}{\includegraphics{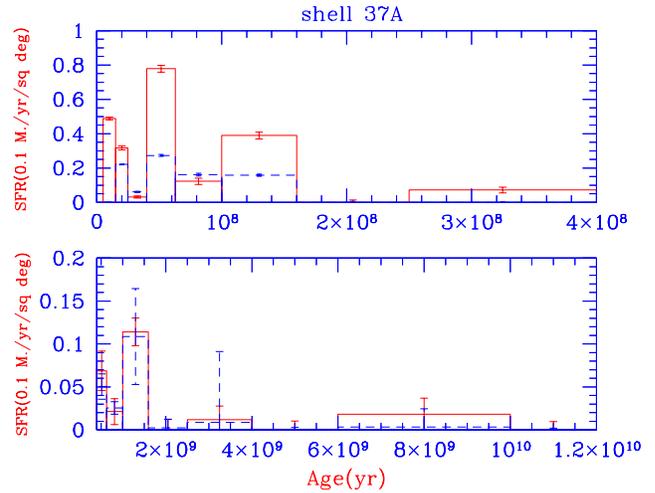}}
\caption{SFR on the East side of the supershell 37A ( solid line ) and on the West side ( dashed line ). }
    \label{shell37A.fig}
    \end{figure}

Analogously to Sections \ref{37A}, \ref{304A} we subdivide the shells in two regions.
Fig. \ref{shell37A.fig} shows the SFR in the shell 37A.
As noticed in the cluster age distribution (see Section \ref{age_det}) the star formation process has been much more active on the East side of  37A, in the region between the two shells.  Concerning the young population, enhancements of the SFR are found at 5, 50, 130 Myr  both on the East and  on the West side. 

\begin{figure}
   \centering	    
\resizebox{8.9cm}{!}{\includegraphics{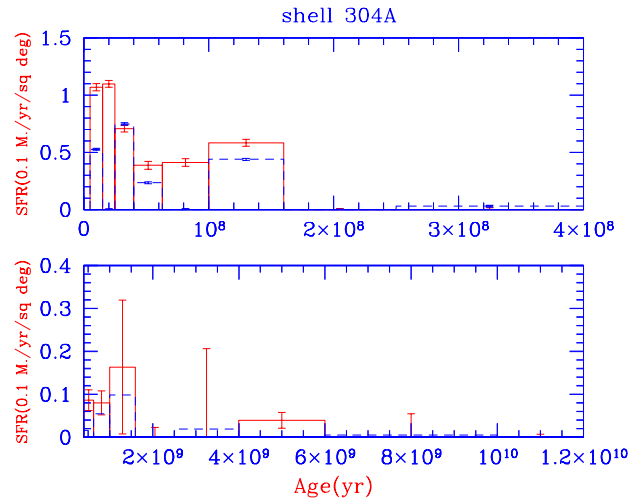}}
\caption{SFR on the Northern side of the supershell 304A ( solid line ) and on the southern side ( dashed line ). }
    \label{shell304A.fig}
    \end{figure}

Fig. \ref{shell304A.fig} presents the SFR in the shell 304A.
 In  the Northern side of the supershell the SF was more active at young ages.  The SF was continuous from a few Myr to about 160 Myr with period of enhancements  from 5 to 40 Myr and between 100-160 Myr. The SFR in both shells is consistent, even if the intensity of the younger episodes was higher in the shell 304A.

Summarizing, 
in both shells the field star formation was continuous in the past 160 Myr. A global burst of SF is found at ages of a few Myr, which might be responsible of the fractal structure of HI interstellar medium \citep{stanimirovic1999, hatzidimitriou2005}.
The SF rate at older ages was less active. Enhancements are found between 100-150 Myr and between 1 and 1.6 Gyr, corresponding to a close interaction between \object{SMC} and \object{LMC}. We remind that the data do not allow precise determination of the SF at older ages.

Comparing with the cluster age distribution (see previous Sections), 
we find that there is not a complete coincidence  between cluster and field star formation, suggesting a different mode of formation.
However,  formation episodes involving both happened at 5, 20, and finally at 100-150  Myr, in coincidence with \object{SMC} perigalactic passage.

At present, no detailed studies of the cluster and star formation due to tidal interaction in the \object{SMC} are available. However
 the evolution and star formation history of \object{LMC} has been derived in great detail by   \citet{bekki2005}. They pointed out that gravitational interactions do not necessarily influence in the same way cluster and field star formation. In fact,
cluster formation is expected to take place only if the perturbation due to tidal effects is strong enough to trigger high velocities   cloud-cloud collision.
 Field star formation is more sensitive to tidal triggers and it is  expected
 to take place under less restrictive conditions. The model by \citet{bekki2005} predicts no cluster formation in the \object{LMC} at the time of the first perigalactic passage, 6.8 Gyr ago, when field star formation was enhanced, while  a strong enhancement of the cluster and field star formation rate in the \object{LMC}  is expected as a consequence of the strong Galaxy-\object{LMC}-\object{SMC} interaction between 2-3.5 Gyr ago. At that time, the peak of the cluster formation is almost coincident with the peak of the field star formation, although slightly delayed, taking place   2.5 Gyr ago. Cluster and field star formation is expected in the \object{LMC}
100-200 Myr ago, at the time of the most recent collision with the \object{SMC}. 

Our results point out  that the  last interaction between \object{SMC} and \object{LMC} has triggered cluster and field star formation in the \object{SMC}, in agreement with what is found in the \object{LMC} and is predicted  by the models.

\section{Summary and conclusions}
\label{conclusions}

In this paper we revise the cluster and field star formation in the main body of the  Small Magellanic Cloud.
 The main goal is to perform a study of the correlation between young objects and their  interstellar environment. The age of  311 clusters and 164 associations is determined through isochrone fitting method.  The spatial distribution of the clusters is compared with the HI maps, with the HI velocity dispersion field,  with the location of the CO clouds and with the distribution of young field stars.

Our main results are as follows:

1) The cluster age distribution supports the idea that clusters formed in the last 1 Gyr of the \object{SMC} history  in a roughly continuous way with periods of enhancement. 
The age distribution of the clusters in the whole disk presents enhancements, namely between  a few Myr and 15 Myr and at 90 Myr.
Old objects are mainly located close to the \object{SMC} center and on the SW side. 
Models of the interactions between \object{LMC}-\object{SMC} and Milky Way
predict a close encounter between the MCs  roughly 100--200 Myr ago. At that time the star formation is expected to be enhanced not only in the tidal arms, but also in the main body of the \object{SMC}.
 In fact an episode at 90 Myr is found in the age distribution of the clusters that might be due to  tidal trigger.
However the age distribution presents younger episodes that might have different origin and are possibly due to local phenomena.

2) The two shells 37A and 304A  are clearly visible in the spatial age distribution of the clusters younger than 15 Myr about: the mechanism responsible of the shell formation (SN, stellar winds, and/or turbulence)  is closely related to the cluster formation.   The regions have been very active especially at the edges of the shells and in the inter-shell region since 1 Gyr ago. 
In the supershell 37A  clusters younger than a few   $10^7$yr are located at the Eastern  rim of the supershell 37A where gas and dust are located, while older episodes are widely distributed. The cluster age distribution at the Eastern part shows a young episode at a few Myr, and several enhancements, namely between a few Myr and 15 Myr, and at 80 Myr. 
 On the Western side, the star formation was less efficient at ages younger than 15 Myr, while it was comparable at older ages. 

The cluster distribution in the supershell 304A shows a continuous formation from a few Myr to 1 Gyr.
The dominant episode was between a few Myr and 20 Myr. An enhancement is found at  90 Myr.

3) We find that star clusters and associations form in  clustered distribution.  The typical correlation scale of the clusters is of the order of 500pc, comparable with the dimensions of the large molecular cloud complexes found in the \object{LMC} and in the \object{SMC}. The two point autocorrelation function of the young massive field stars shows a stronger correlation, but on a comparable scale. 

4) A tight  cross-correlation between young clusters and the HI intensity  is found. The degree of correlation is decreasing with the age of the clusters. Finally clusters older than 300 Myr are located away from the HI peaks.  Clusters and associations younger than 10 Myr are
related to the CO clouds in the SW region of the disk, but not in the NE where smaller clouds are found. Older generation is more evenly distributed.
This is in agreement with the evolutionary time scale of the giant molecular clouds that is found to be of the order of 10 Myr. 
This correlation indicates that the molecular gas content is associated to the field and cluster formation, but that its presence does not necessarily imply star formation. 

5) A weak relation between the location of the young clusters and the velocity dispersion field of the atomic gas is derived. The shell 304A (where a positive correlation is found) is coincident with a high velocity dispersion region where shocks among clouds might have triggered cluster formation. However this mechanism cannot account for  the majority of young objects in the southern shell, 37A where young clusters are located in region of intermediate velocity dispersion.

6) 
The field star formation was continuous in the past 160 Myr. Then periods of quiescence were followed by enhanced activity. A global burst of SF is found at ages of a few Myr, which might be responsible of the fractal structure of HI interstellar medium.
 Enhancements are found between 100-150 Myr and between 1 and 1.6 Gyr, corresponding to a close interaction between \object{SMC} and \object{LMC}. 
The last tidal interaction between the MCs (100-200 Myr ago) has triggered the formation of both clusters and field stars. However, clusters and field  formation rates are not completely coincident, suggesting a  different mode of formation.

\acknowledgements{Many thanks are due to G. Bertelli for many helpfull discussions. This work has been partially supported by INAF PRIN 2002}

\bibliographystyle{aa}

\Online
\clearpage
\onecolumn 

\hoffset=-10mm




\end{table*}

\end{document}